# FINITE DIMENSIONAL REPRESENTATIONS OF QUANTUM AFFINE ALGEBRAS AT ROOTS OF UNITY

Jonathan Beck[1] and Victor G. Kac[2]

**Abstract.** We describe explicitly the canonical map $\chi : \operatorname{Spec} U_\varepsilon(\widetilde{\mathfrak{g}}) \to \operatorname{Spec} Z_\varepsilon$, where $U_\varepsilon(\widetilde{\mathfrak{g}})$ is a quantum loop algebra at an odd root of unity $\varepsilon$. Here $Z_\varepsilon$ is the center of $U_\varepsilon(\widetilde{\mathfrak{g}})$ and Spec $R$ stands for the set of all finite–dimensional irreducible representations of an algebra $R$. We show that Spec $Z_\varepsilon$ is a Poisson proalgebraic group which is essentially the group of points of $G$ over the regular adeles concentrated at 0 and $\infty$. Our main result is that the image under $\chi$ of Spec $U_\varepsilon(\widetilde{\mathfrak{g}})$ is the subgroup of principal adeles.

**Contents.**

## §0. Introduction.

The purpose of this paper is to study finite–dimensional irreducible representations of the quantum loop algebra $\widetilde{U}_\varepsilon = U_\varepsilon(\widetilde{\mathfrak{g}})$ at an odd root of unity $\varepsilon$. Here $\mathfrak{g}$ is a simple finite–dimensional Lie algebra over $\mathbb{C}$ and $\widetilde{\mathfrak{g}} = \mathbb{C}[t, t^{-1}] \otimes_\mathbb{C} \mathfrak{g}$ is the associated loop algebra.

Denoting by Spec $R$ the set of all finite–dimensional irreducible complex representations of an associative algebra $R$ over $\mathbb{C}$ and by $Z$ the center of $R$, we have (by Schur's lemma) the canonical map:

$$(0.1) \qquad \chi : \operatorname{Spec} R \to \operatorname{Spec} Z.$$

(Recall that the value of $\chi$ on a representation $\sigma \in \operatorname{Spec} R$ is defined by $\sigma(z) = \chi(\sigma)I$ for $z \in Z$.) If $R$ is a finitely generated module over $Z$ (which is the case for $R = U_\varepsilon(\mathfrak{g})$ [DC–K]) one knows that the map $\chi$ is surjective with finite fibers and, moreover, it is bijective over a Zariski open dense subset of Spec $Z$. In other words, at least "generically", Spec $Z$ parametrizes the set of all irreducible finite–dimensional irreducible representations of $R$. This well–known observation was the starting point for a thorough (albeit incomplete) study of Spec $U_\varepsilon(\mathfrak{g})$ taken up in [DC–K], [DC–K–P1,2,3] and other papers.

---

[1] Supported by an NSF Postdoctoral Fellowship. [2]Supported in part by NSF grant DMS–9103792.

Typeset by $\mathcal{A}_\mathcal{M}\mathcal{S}$-TeX

In the case when $R = \widetilde{U}_\varepsilon$ the situation is quite different since $\widetilde{U}_\varepsilon$ is not finitely generated over its center $Z = Z_\varepsilon$. The canonical map $\chi$ is not surjective and is not generically bijective. The main result of the present paper is the calculation of the image of $\chi$ in Spec $Z_\varepsilon$.

The first result (Proposition 2.3) provides a (infinite) set of generators of the algebra $Z_\varepsilon$. By general principles, $Z_\varepsilon$ has a canonical structure of a Poisson algebra. Furthermore, we show that $Z_\varepsilon$ is a Hopf subalgebra of the Hopf algebra $\widetilde{U}_\varepsilon$. (Recall that this isn't the case for $U_\varepsilon(\mathfrak{g})$.) Thus, $Z_\varepsilon$ is a Poisson Hopf algebra, and using a "Frobenius homomorphism" we obtain that it is isomorphic to a certain Poisson Hopf algebra $\overline{U}_1$ independent of the odd root of unity $\varepsilon$ (Corollary 3.2.1 and 3.2.2).

In the dual language, Spec $Z_\varepsilon$ is a Poisson proalgebraic group. Our first key result (Theorem 5.3) is the construction of a Poisson group isomorphism

$$(0.2) \qquad \pi : \text{Spec } Z_\varepsilon \to \Omega$$

with a Poisson proalgebraic group $\Omega$ described below. This result and its proof are similar to that in the "finite type" case given by [DC–K–P].

The group $\Omega$ is defined as follows. Let $\underline{G}$ be the connected simply connected algebraic group over $\mathbb{C}$ whose Lie algebra is $\mathfrak{g}$. Consider the triangular decomposition $\mathfrak{g} = \mathfrak{n}_- \oplus \mathfrak{h} \oplus \mathfrak{n}_+$ and let $\underline{N}_\pm$ and $\underline{H}$ be the closed algebraic subgroups of $\underline{G}$ with the Lie subalgebras $\mathfrak{n}_\pm$ and $\mathfrak{h}$ respectively. We denote by $\Omega$ the subgroup of the proalgebraic group $\widetilde{\underline{G}} = \underline{G}(\mathbb{C}[[t^{-1}]]) \times \underline{G}(\mathbb{C}[[t]])$ consisting of elements of the form $(hu_-(t^{-1}), h^{-1}u_+(t))$ where $h \in \underline{H}(\mathbb{C})$, $u_\pm(t^\pm) \in \underline{G}(\mathbb{C}[[t^\pm]])$ and $u_+(0) \in \underline{N}_+(\mathbb{C}), u_-(\infty) \in \underline{N}_-(\mathbb{C})$. The Poisson structure on $\Omega$ is defined by making use of a suitable Manin triple (as explained in §4.1). Here we note only that the symplectic leaves of this Poisson structure on $\Omega$ are connected components of the intersections of $\Omega$ with the orbits of the group $\underline{G}(\mathbb{C}[t, t^{-1}]) \times \underline{G}(\mathbb{C}[t, t^{-1}])$ acting on $\underline{G}(\mathbb{C}((t^{-1}))) \times \underline{G}(\mathbb{C}((t)))$ by $(k_1, k_2) \cdot (a, b) = (k_1 a k_2^{-1}, k_1 b k_2^{-1})$.

In order to describe the image $\mathcal{F}$ of the map $\pi$ in $\Omega$, introduce the following notation. Let $\mathbb{C}(t)_0$ be the subalgebra of the field of rational functions in the indeterminant $t$ consisting of functions regular at 0 and at $\infty$. (This is a semilocal algebra.) We have an embedding $\mathbb{C}(t)_0 \hookrightarrow \mathbb{C}[[t^{-1}]] \times \mathbb{C}[[t]]$ by taking the power series expansions at $\infty$ and 0. Our second key result (Theorem 6.6) is that

$$(0.3) \qquad \mathcal{F} = \Omega \cap \{(g, g) \in \widetilde{\underline{G}} \mid \text{Ad } g \in (\text{Ad } G)(\mathbb{C}(t)_0)\}.$$

In other words $\mathcal{F}$ is described as follows. Let $\mathcal{O}$ be the algebra of algebraic functions in $t$ which are regular at 0 and $\infty$. Consider $g \in \underline{G}(\mathcal{O})$ such that Ad $g$ is defined over $\mathbb{C}(t)_0 \subset \mathcal{O}$, $g(0) \in \underline{HN}_+(\mathbb{C})$, $g(\infty) \in \underline{HN}_-(\mathbb{C})$ and the product of projections of $g(0)$ and $g(\infty)$ on $\underline{H}(\mathbb{C})$ equals 1. Consider the pair $(a, b) \in \underline{G}(\mathbb{C}[[t^{-1}]]) \times \underline{G}(\mathbb{C}[[t]])$ where $a$ (resp. $b$) is the power series expansion of $g$ at $\infty$ (resp. 0). Then $\mathcal{F}$ consists of all such pairs.

We prove (0.3) in two steps. First, we develop a theory of "diagonal" finite–dimensional irreducible representations of $\widetilde{U}_\varepsilon$. A representation $\sigma$ is called *diagonal* if $\chi(\sigma) \in \Omega \cap (\underline{H}(\mathbb{C}[[t^{-1}]]) \times \underline{H}(\mathbb{C}[[t]]))$. We show that these representations are classifed by their "highest weights", which are, essentially, $n$–tuples ($n = \text{rank } \mathfrak{g}$) of rational functions $(R_1(t), \ldots, R_n(t))$ which are regular at 0 and at $\infty$ and such that $R_i(0)R_i(\infty) = 1$ for all $i$ (Theorem 6.3).

Note that any finite–dimensional representation of $U_q(\widetilde{\mathfrak{g}})$ defined for generic $q$ is diagonal when specialized to $q = \varepsilon$. Finite–dimensional irreducible representations of $U_q(\widetilde{\mathfrak{g}})$ were classified in [CP2] by rational functions of a very special form, in agreement with our results.

The second step consists of two parts. First we show that the elements of $\widetilde{U}_\varepsilon$ act "quasipolynomially" in a finite–dimensional representation, which implies the inclusion $\subset$ in (0.3). The reverse inclusion on the "diagonal" part follows from Theorem 6.3. To establish it on the "off diagonal" part we use the fact that along a symplectic leaf of $\Omega$ the representation theory of $\widetilde{U}_\varepsilon$ remains unchanged (cf. [DC–K–P2]).

We do not know a complete classification of finite–dimensional irreducible representations, even in the case of $U_\varepsilon(\widetilde{\mathfrak{sl}_2})$. It follows from our results that we get all central characters by considering irreducible subquotients of tensor products of evaluation representations. Finite–dimensional irreducible representations of $U_\varepsilon(\widetilde{\mathfrak{sl}_n})$ were studied in [T], but we do not understand its connection to our work.

Throughout the paper we denote by $R^\times$ the set of invertible elements of a ring $R$, with the exception that $\mathbb{Z}^\times = \mathbb{Z} \setminus \{0\}$. We denote by $\mathbb{Z}_+$ the set of non–negative integers.

We would like to thank I. Grojnowski and L. Vaserstein for helpful discussions.

## §1. Notation.

1.1 Let $\mathfrak{g}$ be a simple finite–dimensional Lie algebra over $\mathbb{C}$. Choose a Cartan subalgebra $\mathfrak{h}$, let $\Delta \subset \mathfrak{h}^*$ be the set of roots and let $Q = \mathbb{Z}\Delta$ be the root lattice. Let

$$\mathfrak{g} = \mathfrak{h} \oplus (\oplus_{\alpha \in \Delta} \mathfrak{g}_\alpha)$$

be the root space decomposition. For each root $\alpha \in \Delta$ there exists a unique coroot $\alpha^\vee \in [\mathfrak{g}_\alpha, \mathfrak{g}_{-\alpha}] \subset \mathfrak{h}$ such that $<\alpha, \alpha^\vee> = 2$. Let $\Delta^\vee \subset \mathfrak{h}$ be the set of coroots and let $Q^\vee = \mathbb{Z}\Delta^\vee$ be the coroot lattice. Let $P = \{\lambda \in \mathfrak{h}^* | \langle \lambda, Q^\vee \rangle \subset \mathbb{Z}\}$ be the weight lattice and let $P^\vee = \{\lambda \in \mathfrak{h} \mid \langle \lambda, Q \rangle \subset \mathbb{Z}\}$ be the coweight lattice.

Denote by $(.|.)$ the invariant bilinear symmetric form on $\mathfrak{g}$ (and the induced form on $\mathfrak{g}^*$) normalized by the condition that the square length of a short root equals 2. For $\alpha \in \Delta$ let $d_\alpha = \frac{1}{2}(\alpha|\alpha)$—this is a positive integer.

Let $W \subset GL(\mathfrak{h}^*)$ be the Weyl group, i.e. the group generated by reflections $s_\alpha$ ($\alpha \in \Delta$) defined by $s_\alpha(\lambda) = \lambda - <\lambda, \alpha^\vee>\alpha$.

Choose a subset of positive roots $\Delta_+ \subset \Delta$ and let $\Pi = \{\alpha_1, \ldots, \alpha_n\}$ be the set of simple roots. Let $d_i = d_{\alpha_i}$ and let $a_{ij} = <\alpha_i^\vee, \alpha_j> = 2(\alpha_i|\alpha_j)/(\alpha_i|\alpha_i)$ ($i,j = 1, \ldots, n$) be the Cartan integers. Then the $d_i$ are the relatively prime positive integers such that $d_i a_{ij} = d_j a_{ji}$ for all $i,j = 1, \ldots, n$. Let $\omega_1, \ldots, \omega_n \in P$ (resp. $\omega_1^\vee, \ldots, \omega_n^\vee \in P^\vee$) be the fundamental weights (resp. coweights) i.e. $\langle \omega_i, \alpha_j^\vee \rangle = \delta_{ij}$ (resp. $\langle \omega_i^\vee, \alpha_j \rangle = \delta_{ij}$). Note that $\alpha_j = \sum_i a_{ij}\omega_i$ and $\alpha_j^\vee = \sum_i a_{ji}\omega_i^\vee$. We also let $s_i = s_{\alpha_i}$.

Let $\mathfrak{n}_\pm = \oplus_{\alpha \in \Delta_+} \mathfrak{g}_{\pm\alpha}$ be the opposite maximal nilpotent subalgebras of $\mathfrak{g}$. Choose Chevalley generators $e_i \in \mathfrak{g}_{\alpha_i}$ and $f_i \in \mathfrak{g}_{-\alpha_i}$ ($i = 1, \ldots, n$) such that $[e_i, f_i] = \delta_{ij}\alpha_i^\vee$.

Let $\underline{G}$ be the connected simply connected algebraic group over $\mathbb{C}$ with the Lie algebra $\mathfrak{g}$. Let $\underline{N}_\pm$ and $\underline{H}$ be the closed algebraic subgroups of $\underline{G}$ with the Lie subalgebras $\mathfrak{n}_\pm$ and $\mathfrak{h}$ respectively. As usual, we denote by $\underline{G}(R)$ the group of points of $\underline{G}$ over a commutative associative ring $R$. We let $G = \underline{G}(\mathbb{C})$, $N_\pm = \underline{N}_\pm(\mathbb{C})$, $H = \underline{H}(\mathbb{C})$.

Let $\mathcal{B}$ be the braid group on generators $T_1, \ldots T_n$ associated to the Weyl group $W$. For $i = 1, \ldots, n$ let:

$$(1.1.1) \qquad t_i = (\exp f_i)(\exp -e_i)(\exp f_i) \in G.$$

One knows that the map $T_i \mapsto t_i$ extends to a homomorphism $\mathcal{B} \to G$ so that via the adjoint representation of $G$ the action of $\mathcal{B}$ on $\mathfrak{g}$ satisfies:

$$(1.1.2) \qquad T_i(\mathfrak{g}_\alpha) = \mathfrak{g}_{s_i(\alpha)}, \quad T_i|_\mathfrak{h} = s_i.$$

1.2 We proceed to define the associated "extended" (= "affine") objects. Let $\widetilde{Q} = Q \oplus \mathbb{Z}\delta$ be a lattice of rank $n+1$ over $\mathbb{Z}$ with the symmetric bilinear form extending $(.|.)$ on $Q$ by $(Q|\delta) = 0$, $(\delta|\delta) = 0$. Let $\theta$ be the highest root in $\Delta_+ \subset \Delta$ and let $d_0 = d_\theta$. Let $\alpha_0 = \delta - \theta$, so that $(\alpha_0|\alpha_0) = 2d_0$. Then the set of *affine simple roots* $\widetilde{\Pi} = \{\alpha_0\} \cup \Pi$ is a $\mathbb{Z}$–basis of the *affine root lattice* $\widetilde{Q}$. Note that the matrix $(a_{ij} = 2(\alpha_i|\alpha_j)/(\alpha_i|\alpha_i))_{i,j=0}^n$ is the extended Cartan matrix of $\mathfrak{g}$. Note also that $d_i a_{ij} = d_j a_{ji}$ for all $i, j = 0, \ldots, n$.

The *affine root system* is the set $\widetilde{\Delta} = \widetilde{\Delta}^{\mathrm{re}} \cup \widetilde{\Delta}^{\mathrm{im}}$, where

$$\widetilde{\Delta}^{\mathrm{re}} = \{\alpha + n\delta \mid \alpha \in \Delta, n \in \mathbb{Z}\}, \quad \widetilde{\Delta}^{\mathrm{im}} = \{n\delta \mid n \in \mathbb{Z}^\times\}.$$

We let $\widetilde{\Delta}_+ = \widetilde{\Delta}_+^{\mathrm{re}} \cup \widetilde{\Delta}_+^{\mathrm{im}}$, where

$$\widetilde{\Delta}_+^{\mathrm{re}} = \{\alpha + n\delta \mid \alpha \in \Delta, \ n \in \mathbb{N}\} \cup \Delta_+, \quad \widetilde{\Delta}_+^{\mathrm{im}} = \{n\delta \mid n \in \mathbb{N}\}.$$

Note that $\widetilde{\Delta}_+ = \widetilde{Q}_+ \cap \widetilde{\Delta}$, where $\widetilde{Q}_+ = \sum_{j=0}^n \mathbb{Z}_+ \alpha_j$.

The action of $W$ on $Q$ is extended to $\widetilde{Q}$ by letting $W(\delta) = \delta$. Define the reflection $s_0$ of $\widetilde{Q}$ by $s_0(\alpha) = s_\theta(\alpha) + \langle \alpha, \theta^\vee \rangle \delta$. The *affine Weyl group* $\widetilde{W}$ is then the subgroup of $GL(\widetilde{Q})$ generated by all $s_i$, $i = 0, \ldots, n$. Recall that $\widetilde{W}$ is a Coxeter group on generators $\{s_0, \ldots, s_n\}$. Let $\mathcal{T}$ denote the group of all permutations $\sigma$ of the index set $\{0, 1, \ldots, n\}$ such that $a_{\sigma(i),\sigma(j)} = a_{ij}$ $(i, j = 1, \ldots, n)$. This group acts by automorphisms of the lattice $\widetilde{Q}$ by $\sigma(\alpha_i) = \alpha_{\sigma(i)}$ which preserve the bilinear form $(.|.)$. Consider the extended affine Weyl group $\widetilde{W}^e = \mathcal{T} \ltimes \widetilde{W}$. The group $P^\vee$ imbeds in $\widetilde{W}^e$ via $\alpha \mapsto t_\alpha$, where

$$t_\alpha(\beta) = \beta - (\beta|\alpha)\delta \ (\beta \in \widetilde{Q}).$$

Recall that $Q^\vee$ then imbeds in $\widetilde{W}$ so that $\widetilde{W} = W \ltimes Q^\vee$.

Let $\widetilde{\mathcal{B}}$ denote the braid group on generators $T_0, \ldots, T_n$ associated to $\widetilde{W}$, and form the extended braid group $\widetilde{\mathcal{B}}^e = \mathcal{T} \ltimes \widetilde{\mathcal{B}}$ in the obvious way. For $\sigma w \in \widetilde{W}^e$ and a reduced expression $w = s_{i_1} \ldots s_{i_k}$ we let $T_{\sigma w} = \sigma T_{i_1} \ldots T_{i_k}$. This is independent of the choice of the reduced expression.

1.3 The "extended" objects are related to the loop algebra and the loop group in the following well–known way (cf. [K]). Let $\mathbb{C}((t))$ denote the field of Laurent series in $t$, and let $\mathbb{C}[[t]]$ and $\mathbb{C}[t, t^{-1}]$ be its subrings of formal power series and of Laurent polynomials. In what follows, $\mathfrak{g}((t))$, $\mathfrak{g}[[t]]$, and $\mathfrak{g}[t, t^{-1}]$ stand for $\mathbb{C}((t)) \otimes_\mathbb{C} \mathfrak{g}$, $\mathbb{C}[[t]] \otimes_\mathbb{C} \mathfrak{g}$, and $\mathbb{C}[t, t^{-1}] \otimes_\mathbb{C} \mathfrak{g}$. Similarly, we denote by $\underline{G}((t))$, $\underline{G}[[t]]$, and $\underline{G}[t, t^{-1}]$ respectively the groups of points of the algebraic group $G$ over $\mathbb{C}((t)), \mathbb{C}[[t]]$, and $\mathbb{C}[t, t^{-1}]$.

We let $\tilde{\mathfrak{g}} = \mathfrak{g}[t, t^{-1}]$, $\widetilde{G} = \underline{G}[t, t^{-1}]$ be the *loop algebra* and the *loop group*. We note that $\mathfrak{g} \cong 1 \otimes \mathfrak{g}$ is a subalgebra of $\tilde{\mathfrak{g}}$ and $G$ is a subgroup of $\widetilde{G}$.

The *root space decomposition* of $\tilde{\mathfrak{g}}$ is defined as follows:

$$\tilde{\mathfrak{g}} = \mathfrak{h} \oplus (\oplus_{\alpha \in \widetilde{\Delta}} \mathfrak{g}_\alpha),$$

where $\mathfrak{g}_{\alpha + k\delta} = t^k \otimes \mathfrak{g}_\alpha$ $(\alpha \in \Delta, k \in \mathbb{Z})$, $\mathfrak{g}_{k\delta} = t^k \otimes \mathfrak{h}$ $(k \in \mathbb{Z}^\times)$.

Choose $e_\theta \in \mathfrak{g}_\theta$ and $e_{-\theta} \in \mathfrak{g}_{-\theta}$ such that $[e_\theta, e_{-\theta}] = -\theta^\vee$, and let $e_0 = t \otimes e_{-\theta} \in \mathfrak{g}_{\alpha_0}$, $f_0 = t^{-1} \otimes e_\theta \in \mathfrak{g}_{-\alpha_0}$. Then $e_i, f_i$ $(i = 0, \ldots, n)$ are the *Chevalley generators* of $\tilde{\mathfrak{g}}$. Along with $\mathfrak{h}$ they satisfy the well–known collection of defining relations [K].

Let $t_0 = (\exp f_0)(\exp -e_0)(\exp f_0) \in \widetilde{G}$. Then (as in the finite–dimensional case) the map $T_i \mapsto t_i$ extends to a homomorphism $\widetilde{\mathcal{B}} \to \widetilde{G}$ so that (1.1.2) holds for all $i = 0, \ldots, n$ [KP].

1.4 Recall the following construction of the set $\widetilde{\Delta}^{\text{re}}_+$ [Be2, Pa]. Fix an element $x \in Q^\vee \subset \widetilde{W}$ such that $\langle x, \alpha_i \rangle > 0$ for all $i = 1, \ldots n$, and fix a reduced expression $x = s_{j_1} \ldots s_{j_d}$ (in the Coxeter group $\widetilde{W}$).

Let $(i_k)_{k \in \mathbb{Z}}$ be the sequence of integers such that $i_k = j_{k \bmod(d)}$. Then the following two important properties hold:

(1) The roots $\beta_k := \begin{cases} s_{i_1} s_{i_2} \ldots s_{i_{k-1}}(\alpha_{i_k}), & k \geq 0, \\ s_{i_0} s_{i_{-1}} \ldots s_{i_{k+1}}(\alpha_{i_k}), & k < 0 \end{cases}$ comprise $\widetilde{\Delta}^{\text{re}}_+$.

(2) Each subsection $s_{i_k} s_{i_{k+1}} \ldots s_{i_{l-1}} s_{i_l}$ for $k < l$ is reduced.

This definition allows a total order $<$ to be defined on the set of positive roots $\widetilde{\Delta}_+$ given by:

(1.4.1) $\qquad \beta_0 < \beta_{-1} < \beta_{-2} < \cdots < r\delta < s\delta < \cdots < \beta_2 < \beta_1$ if $r < s$.

**Remark 1.4.** We give the following example for $\widetilde{U}_q(\widetilde{\mathfrak{sl}}_3)$. Pick $x = 2\rho \in Q^\vee$ where $2\rho = \sum_{\alpha \in \Delta_+} \alpha$. Then a reduced expression of $2\rho$ is given by $(s_0 s_1 s_2 s_1)^2$ and the ordering (1.4.1) has the form:

$$\delta - \alpha_1 - \alpha_2 < \delta - \alpha_2 < 2\delta - \alpha_1 - \alpha_2 < \delta - \alpha_1 < 3\delta - \alpha_1 - \alpha_2 < 2\delta - \alpha_2 < 4\delta - \alpha_1 - \alpha_2$$
$$< 2\delta - \alpha_1 < \cdots < k\delta < \cdots < \delta + \alpha_1 + \alpha_2 < \alpha_2 < \alpha_1 + \alpha_2 < \alpha_1$$

This order is convex in the sense that if $\alpha \in \widetilde{\Delta}^{\text{re}}_+$ and $\beta \in \widetilde{\Delta}_+$ are such that $\alpha + \beta \in \widetilde{\Delta}_+$, then $\alpha < \alpha + \beta < \beta$. ([Pa]). The following elements form a basis of the vector space spanned by the real root spaces of $\widetilde{\mathfrak{g}}$:

(1.4.2) $\qquad e_{\beta_k} = \begin{cases} t_{i_0} \ldots t_{i_{k+1}}(e_{i_k}), & k \leq 0 \\ t_{i_1} t_{i_2} \ldots t_{i_{k-1}}(e_{i_k}), & k > 0 \end{cases}$

$\qquad e_{-\beta_k} = \begin{cases} t_{i_0} \ldots t_{i_{k+1}}(f_{i_k}), & k \leq 0 \\ t_{i_1} t_{i_2} \ldots t_{i_{k-1}}(f_{i_k}), & k > 0 \end{cases}$

We remark that when $\beta_k = \alpha + k\delta$ for $\alpha \in \widetilde{\Delta}^{\text{re}}$ we have $e_{\alpha + k\delta} = t^k \otimes e_\alpha$.

1.5 One defines the *quantum loop algebra* $\widetilde{U}_q = U_q(\widetilde{\mathfrak{g}})$ of Drinfel'd and Jimbo as an algebra over $\mathbb{C}(q)$ on generators $E_i$, $F_i$ ($i = 0, \ldots, n$), and $K_\alpha$ ($\alpha \in P$), subject to the following relations:

(1.5.1) $\qquad [K_\alpha, K_\beta] = 0, \ K_\alpha K_\beta = K_{\alpha + \beta}, \ K_0 = 1,$

$\qquad K_\alpha E_j K_\alpha^{-1} = q^{(\alpha|\alpha_j)} E_j, \ K_\alpha F_j K_\alpha^{-1} = q^{-(\alpha|\alpha_j)} F_j,$

$\qquad [E_i, F_j] = \delta_{ij} \dfrac{K_i - K_i^{-1}}{q^{d_i} - q^{-d_i}},$

$\qquad (\text{ad}_q E_i)^{1 - a_{ij}}(E_j) = 0, \quad (\text{ad}_q F_i)^{1 - a_{ij}}(F_j) = 0 \ \text{ if } i \neq j.$

We make frequent use of the abbreviated notation $K_i = K_{\alpha_i}$, $[s]_d = \frac{q^{ds} - q^{-ds}}{q^d - q^{-d}}$, $[s]_d! = [1]_d \ldots [s]_d$, $\begin{bmatrix} m \\ r \end{bmatrix}_d = \frac{[m]_d!}{[r]_d! [m-r]_d!}$, and we write $[s]$ for $[s]_1$. The notation $\mathrm{ad}_q$ is explained below (see (1.5.2)).

We recall that $\widetilde{U}_q$ has a Hopf algebra structure with comultiplication $\Delta$, antipode $S$ and counit $\eta$ defined by:

$$\Delta E_i = E_i \otimes 1 + K_{\alpha_i} \otimes E_i, \ \Delta F_i = F_i \otimes K_{-\alpha_i} + 1 \otimes F_i, \ \Delta K_\alpha = K_\alpha \otimes K_\alpha,$$
$$SE_i = -K_{-\alpha_i} E_i, \ SF_i = -F_i K_i, \ SK_\alpha = K_{-\alpha},$$
$$\eta E_i = 0, \ \eta F_i = 0, \ \eta K_\alpha = 1.$$

Then in (1.5.1) and in what follows we define $\mathrm{ad}_q$ by:

(1.5.2) $$(\mathrm{ad}_q x)(y) = \sum_i a_i y S(b_i) \text{ if } \Delta x = \sum_i a_i \otimes b_i.$$

Introduce the $\mathbb{C}$–algebra anti-automorphism $\kappa$ of $U_q$, defined by:

$$\kappa(E_i) = F_i, \quad \kappa(F_i) = E_i, \quad \kappa(K_\alpha) = K_{-\alpha}, \quad \kappa(q) = q^{-1}.$$

Recall that the braid group $\widetilde{\mathcal{B}}$ acts as a group of automorphisms of the algebra $\widetilde{U}_q$ by the following formulas [L]:

$$T_i E_i = -F_i K_i, \ T_i(E_j) = \frac{(-1)^{a_{ij}}}{[a_{ij}]_{d_i}!} (\mathrm{ad}_q E_i)^{-a_{ij}}(E_j) \text{ if } i \neq j,$$
$$T_i K_\alpha = K_{s_i \alpha} \quad (\alpha \in P), \quad \kappa T_i = T_i \kappa.$$

This action is extended to the action of $\widetilde{\mathcal{B}}^e$ in the obvious way.

Let $\widetilde{U}_q^+$ (resp. $\widetilde{U}_q^-$) denote the subalgebra of $\widetilde{U}_q$ generated by the $E_i$ (resp. $F_i$), and let $U_q^0$ be the subalgebra generated by the $K_\alpha$ ($\alpha \in P$). Then multiplication defines an isomorphism ([Ro, L]):

(1.5.3) $$\widetilde{U}_q^- \otimes U_q^0 \otimes \widetilde{U}_q^+ \xrightarrow{\sim} \widetilde{U}_q.$$

Define the subalgebras $\widetilde{U}_q^{\geq 0}$ (resp. $\widetilde{U}_q^{\leq 0}$) generated by $U_q^0$ and the $E_i$ (resp. $F_i$). The algebras $\widetilde{U}_q^+$ and $\widetilde{U}_q^{\geq 0}$ are graded by $\tilde{Q}_+$ in the usual way:

$$\widetilde{U}_q^{+ \text{ (resp.} \geq 0)} = \oplus_\nu (\widetilde{U}_q^{+ \text{ (resp.} \geq 0)})_\nu.$$

1.6 For each $\beta_k \in \widetilde{\Delta}_+^{\mathrm{re}}$ define the *root vector* $E_{\beta_k}$ by:

(1.6.1) $$E_{\beta_k} = \begin{cases} T_{i_0}^{-1} \ldots T_{i_{k+1}}^{-1} (E_{i_k}) & \text{if } k \leq 0 \\ T_{i_1} T_{i_2} \ldots T_{i_{k-1}} T_{i_k} (F_{i_k}) & \text{if } k > 0 \end{cases}$$

**Remark 1.6.** A useful property of these real root vectors is that each (up to a factor from $U_q^0$) is some integral power of $T_x$ ($x = s_{i_0} s_{i_1} \ldots s_{i_d} \in Q^\vee$) applied to the finite set $S = \{T_{i_d}^{-1} \ldots T_{i_{k+1}}^{-1} E_{i_k} \mid 0 \leq k \leq d\}$.

**Definition 1.6.** *For $i = 1, \ldots, n$ and $m > 0$ let*

$$\psi_m^{(i)} = K_i^{-1}[E_i, E_{m\delta - \alpha_i}], \tag{1.6.2}$$

$$\psi_{-m}^{(i)} = \kappa(\psi_m^{(i)}), \quad \psi_0^{(i)} = \frac{K_i - K_i^{-1}}{q^{d_i} - q^{-d_i}}. \tag{1.6.3}$$

For $k > 0$, define *imaginary root vectors* $E_{k\delta}^{(i)}$, $(i = 1, \ldots, n)$ by the following functional equation involving the $\psi_k^{(i)}$:

$$\exp\left((q^{d_i} - q^{-d_i}) \sum_{k=1}^{\infty} E_{k\delta}^{(i)} u^k\right) = 1 + (q^{d_i} - q^{-d_i}) \sum_{k=1}^{\infty} \psi_k^{(i)} u^k. \tag{1.6.4}$$

As usual we extend these definitions to $\widetilde{U}_q^{\leq 0}$ using the antiautomorphism $\kappa$: $E_{-\beta} := \kappa(E_\beta)$ for $\beta \in \widetilde{\Delta}_+$.

These root vectors have the nice property that up to a sign they coincide with the Drinfel'd generators [D]. Namely, we have:

**Theorem 1.6** [Be]. *The algebra $\widetilde{U}_q$ is an associative algebra on generators ($i = 1, \ldots, n$, $k \in \mathbb{Z}$):*

$$E_\beta \ (\beta = \pm \alpha_i + k\delta \in \widetilde{\Delta}), \ E_{k\delta}^{(i)} \ (k \neq 0), \ K_\alpha \ (\alpha \in P),$$

*and the following relations:*

(1.6.5a) $[K_\alpha, E_{k\delta}^{(i)}] = [K_\alpha, K_\beta] = 0$, $K_\alpha E_\beta K_\alpha^{-1} = q^{(\alpha|\beta)} E_\beta$,

(1.6.5b) $[E_{k\delta}^{(i)}, E_{l\delta}^{(j)}] = 0$, $[E_{k\delta}^{(i)}, E_{\pm\alpha_j + l\delta}] = \pm \varepsilon_{ij}^{kl} \frac{1}{k}[ka_{ij}]_{d_i} E_{\pm\alpha_j + (k+l)\delta}$,

(1.6.5c) $E_{\pm\alpha_i + (k+1)\delta} E_{\pm\alpha_j + l\delta} - q^{\pm(\alpha_i|\alpha_j)} E_{\pm\alpha_j + l\delta} E_{\pm\alpha_i + (k+1)\delta}$
$= \varepsilon_{ij}^{kl}(q^{\pm(\alpha_i|\alpha_j)} E_{\pm\alpha_j + (l+1)\delta} E_{\pm\alpha_i + k\delta} - E_{\pm\alpha_i + k\delta} E_{\pm\alpha_j + (l+1)\delta})$,

(1.6.5d) $[E_{\alpha_i + k\delta}, E_{-\alpha_j + l\delta}] = \delta_{ij} K_i^{\text{sgn}(k+l)} \psi_{k+l}^{(i)}$,

(1.6.5e) $\text{Sym}_{k_1, k_2, \ldots, k_m} \sum_{r=0}^{m} (-1)^r \varepsilon_{ij}^{\vec{k}l} \begin{bmatrix} m \\ r \end{bmatrix}_{d_i}$
$E_{\pm\alpha_i + k_1\delta} \ldots E_{\pm\alpha_i + k_r\delta} E_{\pm\alpha_j + l\delta} E_{\pm\alpha_i + k_{r+1}\delta} \ldots E_{\pm\alpha_i + k_m\delta} = 0$,

*where $i \neq j$, $m = 1 - a_{ij}$, and $\varepsilon_{ij}^{kl} = \pm 1$ as explained in [Be]. Sym denotes symmetrization with respect to the indices $\vec{k} = (k_1, k_2, \ldots k_m)$. The function $\text{sgn}(x)$ is defined to be $\frac{|x|}{x}$ for $x \neq 0$ and $\text{sgn}(0) = 0$.* □

1.7 As defined, the root vectors $E_\beta$ ($\beta \in \widetilde{\Delta}_+$) are in $\widetilde{U}_q^{\geq 0}$. Defining $\dot{E}_{\alpha + k\delta} = E_{\alpha + k\delta}$ (resp. $= K_{-\alpha} E_{-\alpha + k\delta}$) if $\alpha \in \Delta_+$ and $\dot{E}_{k\delta} = E_{k\delta}$, we have $\dot{E}_\beta \in \widetilde{U}_q^+$ for all $\beta \in \widetilde{\Delta}_+$. Introduce the monomials $M_{(a_\beta)} = \Pi_{\widetilde{\Delta}_+, <} \dot{E}_\beta^{a_\beta} \in \widetilde{U}_q^+$. Here $(a_\beta) \in \mathbb{Z}_+^{\widetilde{\Delta}_+}$, where $\mathbb{Z}_+^{\widetilde{\Delta}_+}$ denotes the set of maps $f: \widetilde{\Delta}_+ \mapsto \mathbb{Z}_+$ with finite support and $<$ denotes that the product is convexly ordered. Let $N_{(a_\beta)} := \kappa(M_{(a_\beta)}) \in \widetilde{U}_q^-$.

**Proposition 1.7.** [Be2] (a) *The $M_{(a_\beta)}$ form a basis of $\widetilde{U}_q^+$ over $\mathbb{C}(q)$.*

(b) *The elements $N_{(a_\beta)} K_\alpha M_{(a'_\beta)}$, where $(a_\beta), (a'_\beta) \in \mathbb{Z}_+^{\tilde{\Delta}_+}$, $\alpha \in P$, form a basis of $\widetilde{U}_q$ over $\mathbb{C}(q)$.*

(c) *Let $\alpha, \beta \in \tilde{\Delta}_+$ be such that $\beta > \alpha$. Then:*

$$\dot{E}_\beta \dot{E}_\alpha - q^{(\alpha|\beta)} \dot{E}_\alpha \dot{E}_\beta = \sum_{\alpha < \gamma_1 < \cdots < \gamma_n < \beta} c_{\vec{\gamma}} \dot{E}_{\gamma_1}^{a_1} \ldots \dot{E}_{\gamma_n}^{a_n}.$$

*where $c_{\vec{\gamma}} \in \mathbb{C}[q, q^{-1}]$ for $\vec{\gamma} = (\gamma_1, \gamma_2, \ldots, \gamma_n), \in \tilde{\Delta}_+^n$.* □

1.8 Using the above PBW type basis we define a filtration on $\widetilde{U}_q$ as in [DC–K]. Consider a monomial $N_{(a_\beta)} K_\alpha M_{(a'_\beta)}$ where $a_\beta, a'_\beta \in \mathbb{Z}_+^{\tilde{\Delta}_+}$ and $\alpha \in P$. Define its total height by

$$d_0(N_{(a_\beta)} K_\alpha M_{(a'_\beta)}) = \sum_\beta (a_\beta + a'_\beta) \mathrm{ht}\, \beta,$$

and its total degree by

$$d(N_{(a_\beta)} K_\alpha M_{(a'_\beta)}) = (d_0(N_{(a_\beta)} K_\alpha M_{(a'_\beta)}), (a_\beta), (a'_\beta)) \in \mathbb{Z}^{2\tilde{\Delta}_+ + 1}.$$

We view $\mathbb{Z}_+^{2\tilde{\Delta}_+ + 1}$ as a totally ordered semigroup with the usual lexicographical order.

Introduce a $\mathbb{Z}_+^{2\tilde{\Delta}_+ + 1}$–filtration of the algebra $\widetilde{U}_q$ by letting $U_s$ ($s \in \mathbb{Z}_+^{2\tilde{\Delta}_+ + 1}$) be the span of the monomials $N_{(a_\beta)} K_\alpha M_{(a'_\beta)}$ such that $d(N_{(a_\beta)} K_\alpha M_{(a'_\beta)}) \leq s$. Proposition 1.7 implies:

**Proposition 1.8.** *The associated graded algebra $Gr\, \widetilde{U}_q$ of the $\mathbb{Z}_+^{2\tilde{\Delta}_+ + 1}$–filtered algebra $\widetilde{U}_q$ is an algebra over $\mathbb{C}(q)$ on generators $E_\alpha$ ($\alpha \in \tilde{\Delta}_+$ counting multiplicities) and $K_\beta$ ($\beta \in P$) subject to the following relations:*

(1.8.1)
$$K_\alpha K_\beta = K_{\alpha+\beta},\ K_0 = 1;$$
$$K_\alpha E_\beta = q^{(\alpha|\beta)} E_\beta K_\alpha,\ K_\alpha F_\beta = q^{-(\alpha|\beta)} F_\beta K_\alpha;$$
$$E_\alpha E_{-\beta} = E_{-\beta} E_\alpha,\ \text{if } \alpha, \beta \in \tilde{\Delta}_+;$$
$$E_\alpha E_\beta = q^{(\alpha|\beta)} E_\beta E_\alpha,\ E_{-\alpha} E_{-\beta} = q^{(\alpha|\beta)} E_{-\beta} E_{-\alpha},\ \text{if } \alpha, \beta \in \tilde{\Delta}_+ \text{ and } \beta < \alpha.$$ □

1.9 Fix a primitive $\ell$–th root of unity $\varepsilon$. Let $\mathcal{A}^\varepsilon$ be the subring of $\mathbb{C}(q)$ consisting of rational functions regular at $q = \varepsilon$. Let $\tilde{U}_{\mathcal{A}^\varepsilon}$ be the $\mathcal{A}^\varepsilon$–subalgebra of $\widetilde{U}_q$ generated by the elements $E_i, F_i, K_i^{\pm 1}$ and $\psi_0^{(i)}$. Let $(q - \varepsilon)\tilde{U}_{\mathcal{A}^\varepsilon}$ be the 2–sided ideal generated by $(q - \varepsilon)$ in $\tilde{U}_{\mathcal{A}^\varepsilon}$. Define the algebra $\widetilde{U}_\varepsilon$ over $\mathbb{C}$, the *specialization* of $\widetilde{U}_q$ at $\varepsilon$, by $\widetilde{U}_\varepsilon = \tilde{U}_{\mathcal{A}^\varepsilon}/(q - \varepsilon)\tilde{U}_{\mathcal{A}^\varepsilon}$.

**Remark 1.9.** The algebra $\widetilde{U}_\varepsilon$ is the associative algebra over $\mathbb{C}$ on generators $E_i, F_i$, ($i = 0, \ldots, n$), $K_\alpha$ ($\alpha \in P$) and defining relations (1.5.1) where $q$ is replaced by $\varepsilon$, provided that $\ell > \max_i(d_i)$. Of course, all relations (1.6.5) with $q$ replaced by $\varepsilon$ hold in $\tilde{U}_\varepsilon$. Also, $Gr\, \widetilde{U}_\varepsilon$ is the algebra over $\mathbb{C}$ obtained by substituting $q$ for $\varepsilon$ in (1.8.1).

1.10 We make explicit the above formulas for $U_q(\widetilde{\mathfrak{sl}_2})$ (cf. [Da]). In this case $\Delta_+ = \{\alpha\}$, $\omega = \frac{1}{2}\alpha$ is the only fundamental weight, and there is a unique choice of the sequence $(i_k)_{k \in \mathbb{Z}}$, namely $i_k = k \pmod 2$ (see [Be2]). Then $\beta_k = \alpha - k\delta$ for $k \leq 0$ and $\beta_k = -\alpha + k\delta$ for $k > 0$. Define the generators $E_{\alpha-k\delta}$, $E_{-\alpha+k\delta}$, $E_{k\delta}$ as in formulas (1.6.2–5). Then $\widetilde{U}_q = U_q(\widetilde{\mathfrak{sl}_2})$ is the algebra over $\mathbb{C}(q)$ on generators $K_\omega^{\pm 1}, E_{\pm\alpha+k\delta}$ ($k \in \mathbb{Z}$), and $E_{k\delta}$ ($k \in \mathbb{Z}^\times$) with defining relations:

(1.10.1)
$$\text{(a) } [K_\omega, E_{k\delta}] = 0, \ K_\omega E_\beta K_\omega^{-1} = q^{(\omega|\beta)} E_\beta \ (\beta = \pm\alpha + k\delta),$$

$$\text{(b) } [E_{k\delta}, E_{\pm\alpha+l\delta}] = \pm \frac{1}{k}[2k] E_{\pm\alpha+(k+l)\delta}, \ [E_{k\delta}, E_{l\delta}] = 0,$$

$$\text{(c) } E_{\pm\alpha+(k+1)\delta} E_{\pm\alpha+l\delta} - q^{\pm 2} E_{\pm\alpha+l\delta} E_{\pm\alpha+(k+1)\delta}$$
$$= q^{\pm 2} E_{\pm\alpha+(l+1)\delta} E_{\pm\alpha+k\delta} - E_{\pm\alpha+k\delta} E_{\pm\alpha+(l+1)\delta},$$

$$\text{(d) } [E_{\alpha+k\delta}, E_{-\alpha+l\delta}] = K_\omega^{2\mathrm{sgn}(k+l)} \psi_{k+l},$$

where $\psi_m = \psi_m^{(1)}$ are defined by (1.6.2–3).

## §2. The center $Z_\varepsilon$ of $\widetilde{U}_\varepsilon$.

2.1 It is clear from (1.6.5b) that before proceeding to the calculation of $Z_\varepsilon$ we need to calculate the quantity $\Delta_k := \det([ka_{ij}]_{d_i})_{i,j=1}^n$.

**Lemma 2.1.** *The determinants $\Delta_k$ are given by the following formulas:*

$$\begin{array}{llll}
A_n: & [k]^{n-1}[(n+1)k] & B_n: & [k]_2^{n-1}[k][2]_{(2n-1)k} \\
C_n: & [k]^{n-1}[k]_2[2]_{(n+1)k} & D_n: & [k]^{n-1}[2k][2]_{(n-1)k} \\
E_6: & [k]^5[3k]([2]_{4k} - 1) & E_7: & [k]^6[2k]([2]_{6k} - 1) \\
E_8: & [k]^8([2]_{8k} + [2]_{6k} - [2]_{2k} - 1) & F_4: & [k]^2[k]_2^2([2]_{6k} - 1) \\
G_2: & [k][3k]_3([2]_{10k} + [2]_{8k} - [2]_{2k} - 1). \ \square
\end{array}$$

**Corollary 2.1.** *Let $q = \varepsilon$ be a primitive $\ell$–th root of $1$ where $\ell$ is odd. Then $\Delta_k \neq 0$ for all non–zero integers $k$ such that $\ell \nmid k$ provided that the following conditions on $\ell$ hold:*

(2.1.1)
$$A_n \text{ and } C_n : \gcd(\ell, n+1) = 1;$$
$$B_n : \gcd(\ell, 2n-1) = 1; \ D_n : \gcd(\ell, n-1) = 1;$$
$$E_6 \text{ and } G_2 : \gcd(\ell, 3) = 1. \ \square$$

**Remark 2.1.** The conditions (2.1.1) on $\ell$ follows in all cases except for $G_2$ from the condition $\gcd(\ell, h^\vee) = 1$, where $h^\vee$ is the dual Coxeter number of $\mathfrak{g}$ [K, Chapter 6].

2.2 From now on we shall assume that $\ell$ is an odd integer greater than 1 satisfying the conditions (2.1.1). Fix a primitive $\ell$–th root of unity $\varepsilon$. We turn now to the calculation of the center $Z_\varepsilon$ of $\widetilde{U}_\varepsilon$.

**Lemma 2.2** (a) *Let $\beta \in \widetilde{\Delta}^{\mathrm{re}}$. Then $E_\beta^\ell \in Z_\varepsilon$ (for any $\ell > \max_i d_i$).*

(b) *Let $i = 1, \ldots, n, \ k \in \mathbb{Z}^\times$. Then $E_{k\ell\delta}^{(i)} \in Z_\varepsilon$ ( for any $\ell > 1$).*

*Proof.* (a) As shown in [DC–K–P2], $(\text{ad}_\varepsilon E_i)^\ell(E_j) = E_i^\ell E_j - E_j E_i^\ell = 0$ by the last of relations (1.5.1). Similarly one checks that the $F_i^\ell$ are central. Now the other real root vectors are braid group translates of the $E_i$ or $F_i$. (b) follows from (1.6.5a,b). □

Since the algebra $\text{Gr}\,\widetilde{U}_\varepsilon$ is quasipolynomial over $\mathbb{C}$, its center $\overline{Z}_\varepsilon$ can be calculated using the methods of [DC–K–P2]. Let $A = (a_{\beta,\beta'})$, where $a_{\beta,\beta'} = -a_{\beta',\beta} = (\beta|\beta')$ if $\beta < \beta' \in \widetilde{\Delta}_+$ and $a_{\beta,\beta} = 0$ (so that $A$ is antisymmetric). Let $B = ((\omega_i|\beta))$ where $i = 1,\ldots,n$ and $\beta \in \widetilde{\Delta}_+$. Form the infinite matrix indexed by the set $M = \widetilde{\Delta} \cup \{1,\ldots,n\}$:

$$S = \begin{pmatrix} A & -{}^tB & 0 \\ B & 0 & -B \\ 0 & {}^tB & -A \end{pmatrix}$$

Its matrix elements are the commutation coefficients of the algebra $\text{Gr}\,\widetilde{U}_\varepsilon$ in the ordered basis given by Proposition 1.7. Consider the range of $S$ mod $\ell$, i.e. $S : \mathbb{Z}^M \to (\mathbb{Z}/\ell)^M$ and let $H_S$ be the kernel of this map. Then as in [DC–K–P2, Proposition 3.3], a basis for $\overline{Z}_\varepsilon$ is given by $\{\prod_< E_\alpha^{h_\alpha} \mid h = (h_\alpha) \in H_S \cap \mathbb{Z}_+^M\}$. Given a basis of $H_S$ we obtain a polynomial basis of $\text{Gr}\,Z_\varepsilon$. Such a basis of $H_S$ can be made apparent (see [DC–K–P2]) by finding the elementary divisors of the matrix

$$S_1 = \begin{pmatrix} 0 & 0 & 0 \\ -A & 0 & {}^tB \\ 0 & 0 & 0 \end{pmatrix}$$

over $\mathbb{Z}[\frac{1}{2}]$.

We bring $S_1$ to a matrix $S'$ which will have the same elementary divisors using the following row operations on $A$ to obtain a matrix $A'$:

(1) $\text{Row}(-\alpha + k\delta) - \text{Row}(-\alpha + (k+1)\delta) \to \text{Row}(-\alpha + k\delta)$, and
(2) $\text{Row}(\alpha + (k-1)\delta) - \text{Row}(\alpha + k\delta) \to \text{Row}(\alpha + (k-1)\delta)$.

Then

$$A' = \begin{pmatrix} T_1 & 0 & 0 \\ 0 & 0 & 0 \\ 0 & 0 & T_2 \end{pmatrix} \quad \text{and} \quad S' = \begin{pmatrix} 0 & 0 & 0 \\ -A' & 0 & {}^tB' \\ 0 & 0 & 0 \end{pmatrix}$$

where $T_1$ is upper triangular and $T_2$ is lower triangular with the diagonal elements of $T_1$ and $T_2$ equal 1. Hence the kernel is generated by $\ell$–th powers of the real root vectors and the imaginary root vectors. Thus, we have proved

**Proposition 2.2.** $\overline{Z}_\varepsilon$ *is generated by* $K_\alpha^\ell$ $(\alpha \in P)$, $E_\beta^\ell$ $(\beta \in \widetilde{\Delta}^{\text{re}})$ *and* $E_{k\delta}^{(i)}$ $(i = 1,\ldots,n, k \in \mathbb{Z}^\times)$. □

2.3 We apply the previous considerations to calculating the center $Z_\varepsilon$ of $U_\varepsilon$. $Z_\varepsilon$ inherits a filtration from $U_\varepsilon$, and it is straightforward that $\text{Gr}\,Z_\varepsilon \subset \overline{Z}_\varepsilon$.

**Lemma 2.3.1.** *Let $\ell$ satisfy the conditions* (2.1.1). *Let $P(E_{k\delta}^{(i)})$ be a polynomial in the $E_{k\delta}^{(i)}$ $(k > 0)$ where $\Delta_k \neq 0$ for some $k$. Then there exists $j$, $1 \leq j \leq n$, for which $[P(E_{k\delta}^{(i)}), E_{-\alpha_j+\delta}] \neq 0$.*

*Proof.* For notational convenience denote by $c_k^{i,j}$ the $q$–coefficient $[ka_{ij}]$ evaluated at $\varepsilon$. Given a monomial $\prod_{m=1}^{D} E_{k_m\delta}^{(i_m)}$ we have the following formula

$$(2.3.1) \quad [\prod_{m=1}^{D} E_{k_m\delta}^{(i_m)}, E_{-\alpha_j+\delta}] = \sum_{\substack{S\subset\{1,\ldots,D\}\\ S\neq\emptyset}} \prod_{m\in S} c_{k_m}^{i,j} \prod_{m'\in S^c} E_{k_{m'}\delta}^{(i_{m'})} E_{-\alpha_j+(\sum_{m\in S} k_m+1)\delta}.$$

which is calculated from (1.6.5).

Write each monomial $\prod_{m=1}^{D} E_{k_m\delta}^{(i_m)}$ in $P$ in non–decreasing order with respect to the $k_m \geq 1$. Without loss of generality assume that each monomial in $P(E_{k\delta}^{(i)})$ has a factor $E_{k\delta}^{(i)}$ such that $\ell \nmid k$. Pick such a monomial for which $D$ is maximal and $k_1$ is minimal. Summing over all monomials with the same $k_1, k_2, \ldots, k_D$ and the same $i_2, \ldots, i_D$ we see that $P(E_{k\delta}^{(i)})$ is of the form

$$(2.3.2) \quad (\sum_{r=1}^{n} a_r E_{k_1\delta}^{(i_r)}) E_{k_2\delta}^{(i_2)} \ldots E_{k_D\delta}^{(i_D)}$$

$$+ \text{ other algebraically independent expressions.}$$

By the assumption of the lemma there exists a $E_{-\alpha_{j_1}+\delta}$ for which $[\sum_r a_r E_{k_1\delta}^{(r)}, E_{-\alpha_{j_1}+\delta}] = \sum_r c_{k_1}^{r,j_1}(E_{-\alpha_{j_1}+(k_1+1)\delta}) \neq 0$ and so

$$[P(E_{k\delta}^{(i)}), E_{-\alpha_{j_1}+\delta}] = \sum_r a_r c_{k_1}^{r,j} \prod_{m=2}^{D} E_{k_m}^{(i_m)} E_{-\alpha_{j_1}+k_1\delta}$$

$$+ \text{ other algebraically independent expressions} \neq 0. \quad \square$$

**Lemma 2.3.2.** *Suppose that $\ell$ is an odd integer greater than $1$ satisfying (2.1.1). Then $\operatorname{Gr} Z_\varepsilon$ is the subalgebra of $\overline{Z}_\varepsilon$ generated by $K_\alpha^\ell, E_\beta^\ell, E_{j\ell\delta}^{(i)}$ for $\alpha \in P, \beta \in \widetilde{\Delta}^{\operatorname{re}}, j \in \mathbb{Z}^\times, i = 1, \ldots, n$.*

*Proof.* Certainly the above generators are in $\operatorname{Gr} Z_\varepsilon$. Take an expression $N_{(a_\beta)} K_\alpha^\ell M_{(a'_\beta)} \in \operatorname{Gr} Z_\varepsilon$ which is not homogeneous with respect to the inherited grading. By assumption we can find $x \in \widetilde{U}_\varepsilon$ so that $d(x) < d(N_{(a_\beta)} K_\alpha^\ell M_{(a'_\beta)})$ and

$$(2.3.3) \quad N_{(a_\beta)} K_\alpha^\ell M_{(a'_\beta)} + x \in Z_\varepsilon.$$

We show that neither $F_{k\delta}^{(i)}$ or $E_{k\delta}^{(i)}$ appears in $N_{(a_\beta)} K_\alpha^\ell M_{(a'_\beta)}$ where $\ell$ does not divide $k$. We can assume without loss of generality that such an $E_{k\delta}^{(i)}$ appears in $M_{(a_\beta)}$. Collect from $x$ all monomials that have the same factors in the PBW basis as $N_{(a_\beta)} K_\alpha^\ell$ and the real root vectors of $M_{(a'_\beta)}$. Then (2.3.3) can be written as

$$(2.3.4) \quad N_{(a_\beta)} K_\alpha^\ell M_{(a'_\beta),\beta\leq 0} P(E_{k\delta}^{(i)}) M_{(a'_\beta),\beta>0} + x'.$$

where $P$ is a polynomial in $E_{k\delta}^{(i)}$ for $i = 1, \ldots, n$, $k > 0$. We also know that

$$(2.3.5) \qquad [N_{(a_\beta)}K_\alpha^\ell M_{(a'_\beta),\beta\leq 0}P(E_{k\delta}^{(i)})M_{(a'_\beta),\beta>0} + x', E_{-\alpha_j}] = 0.$$

where $d(x') < d(N_{(a_\beta)}K_\alpha^\ell M_{(a'_\beta),\beta\leq 0}P(E_{k\delta}^{(i)})M_{(a'_\beta),\beta>0})$, $j = 1\ldots n$. Since $[M_{(a'_\beta),\beta\leq 0}, E_{-\alpha_j}] = [M_{(a'_\beta),\beta>0}, E_{-\alpha_j}] = 0$, using the grading on $N^-$ it follows that $[P(E_{k\delta}^{(i)}), E_{-\alpha_j}] = 0$ for each $j = 1, \ldots, n$. By Lemma 2.3.1 it follows that $\ell|k_m$ for each $E_{k_m\delta}^{(i_m)}$. □

We have shown the following:

**Proposition 2.3.** *Let $\varepsilon$ be a primitive $\ell$-th root of unity, where $\ell$ is as in Lemma 2.3.2. Then the center $Z_\varepsilon$ of $\widetilde{U}_\varepsilon$ is generated by the elements $E_\beta^\ell$ ($\beta \in \widetilde{\Delta}^{\mathrm{re}}$), $E_{j\ell\delta}^{(i)}$ ($j \in \mathbb{Z}^\times, i = 1, \ldots, n$), $K_\alpha^\ell$ ($\alpha \in P$).* □

**Remark 2.3.** Proposition 2.3 also holds for the central extension $U_\varepsilon(\hat{\mathfrak{g}})$ of $U_\varepsilon(\widetilde{\mathfrak{g}})$ (where we add the central element $C$ of $U_\varepsilon(\hat{\mathfrak{g}})$).

### §3. The Frobenius isomorphism and the Poisson structure on $Z_\varepsilon$.

3.1 It is known that $Z_\varepsilon$ is a Poisson algebra as explained in [DC–K–P], a Poisson structure being given by the formula:

$$(3.1.1) \qquad \{x, y\} = \frac{[\tilde{x}, \tilde{y}]}{\ell(q^\ell - q^{-\ell})} \mod (q - \varepsilon).$$

Here $\tilde{x}$ stands for a preimage of $x \in Z_\varepsilon \subset \widetilde{U}_\varepsilon$ under the specialization map.

**Lemma 3.1.** *Let $i = 1, \ldots, n$, then $\{E_i^\ell, E_{\delta-\alpha_i}^\ell\} = K_i^\ell \ell(\varepsilon - \varepsilon^{-1})E_{k\ell\delta}^{(i)}$.*

*Proof.* We consider the case of $\widetilde{\mathfrak{sl}_2}$, the general case follows from this case and (1.6.5c). From §2 we know that $E_\alpha^\ell$, $E_{\delta-\alpha}^\ell$ are central. From Proposition 2.3 we see

$$(3.1.2)\ [E_\alpha^\ell, E_{\delta-\alpha}^\ell] = (q-\varepsilon)\sum c_{(k,k',k'')} \prod_{<,k\in K} E_{\alpha+k\delta}^\ell P(E_{k'\ell\delta}) \prod_{<,k''\in K''} E_{-\alpha+k''\delta}^\ell + (q-\varepsilon)^2 y.$$

Using relation 1.7c) we see that if $k \neq 0$ for for some $k \in K$ then $k'' \neq 0$ for some $k'' \in K''$. This is impossible since $k + k''$ must equal 1. Therefore the bracket (3.1.2) must be purely imaginary and central and therefore, by Prop. 2.3, a multiple of $E_{\ell\delta}$. The coefficient is easily calculated to be $\ell(\varepsilon - \varepsilon^{-1})$ by using (1.10.1d) and (1.6.4). □

3.2 In this section we introduce a renormalized version of $\widetilde{U}_q$ which on specialization to 1 is isomorphic as a Poisson algebra to $Z_\varepsilon$. In the finite type case this variation was considered in the works of [Re] and [DC–P].

**Definition 3.2 (a).** Let $\overline{U}_{\mathcal{A}^1}$ be the subalgebra of $\widetilde{U}_q$ generated by:

$$\overline{E}_\alpha = (q^{d_i} - q^{-d_i})E_\alpha, \quad K_\beta \ (\alpha \in \widetilde{\Delta},\ \beta \in P).$$

**(b)** Let $\overline{U}_1 = \overline{U}_{\mathcal{A}^1}/(q-1)\overline{U}_{\mathcal{A}^1}$.

$\overline{U}_1$ has a Poisson structure given by (cf. (3.1.1)):

$$\{x, y\} = \frac{[\tilde{x}, \tilde{y}]}{(q - q^{-1})} \mod (q - 1).$$

**Lemma 3.2.1.** $\overline{U}_1$ *is a commutative algebra. As a Poisson algebra* $\overline{U}_1$ *is generated by* $\overline{E}_i, \overline{F}_i, K_\alpha, i = 0, \ldots, n, \ \alpha \in P$.

*Proof.* See [DC–K–P], [DC–P]. $\square$

**Lemma 3.2.2.** *There exists a unique Poisson algebra homomorphism* $Fr : \overline{U}_1 \to Z_\varepsilon$ *such that* $\overline{E}_\alpha \mapsto E_\alpha^\ell$, *for* $\alpha \in \widetilde{\Delta}^{\mathrm{re}}$.

*Proof.* $\{\overline{E}_i | \ i = 0, 1, \ldots, n\}$ form a set of Poisson algebra generators for $\overline{U}_1^+$ with the defining relations $\mathrm{ad}_P^{1-a_{ij}} \overline{E}_i(\overline{E}_j) = 0$, where $\mathrm{ad}_P$ denotes the Poisson adjoint action. Let $Fr$ be defined as in the lemma. Using [L, Theorem 35.1.8] it follows that $Fr$ is well defined as a Poisson algebra homomorphism. Since as defined this map is compatible with the braid group action [L, 41.1.9], $Fr$ satisfies the conditions of the lemma. Since $\overline{U}_1$ is generated as a Poisson algebra by $E_\alpha, \alpha \in \widetilde{\Delta}^{\mathrm{re}}$, $Fr$ is necessarily unique. $\square$

**Corollary 3.2.1.** $Fr$ *is a Hopf algebra homomorphism.*

*Proof.* It is an easy calculation that $\Delta(E_i^\ell) = K_i^\ell \otimes E_i^\ell + E_i^\ell \otimes 1$, $\Delta(F_i^\ell) = F_i^\ell \otimes K_i^{-\ell} + 1 \otimes F_i^\ell$, and $\Delta(K_i^\ell) = K_i^\ell \otimes K_i^\ell$. Thus $Fr$ and $\Delta$ commute on $\overline{E}_i, \overline{F}_i, K_i$, which are Poisson generators of $\overline{U}_1$. Since coproduct and Poisson bracket commute the statement follows. $\square$

Furthermore we have:

**Lemma 3.2.3.** *Let* $k > 0$. *Then* $Fr(\overline{E}_{k\delta}) = \ell(\varepsilon - \varepsilon^{-1}) E_{k\ell\delta}$.

*Proof.* We have an explicit set of polynomial generators for both algebras and the lemma will follow by calculating the images of the imaginary root vectors under $Fr$. It follows from [CP] that modulo the ideal in $\overline{U}_1$ generated by the real root vectors, $\overline{E}_{k\delta}$ is primitive for each $k \in \mathbb{Z}$. For the same reason (and since the center is closed under coproduct by the previous Corollary) it follows that $E_{k\ell\delta}$ is primitive modulo the $\ell$-th powers of the real root vectors. Since $Fr$ and coproduct commute, it follows that $Fr(\overline{E}_{k\delta})$ is primitive (modulo the real root vectors). Now it follows (as in Lemma 3.1) that $Fr(K\overline{E}_{k\delta}) = Fr(\{E_\alpha, E_{k\delta-\alpha}\}) = K^\ell P(E_{k'\ell\delta})_{k' \leq k}$. Since $P(E_{k'\ell\delta})_{k' \leq k}$ is primitive, it must be a multiple of $E_{k\ell\delta}$. It follows using 1.10.1(d) and 1.6.4 that this coefficient is $\ell(\varepsilon - \varepsilon^{-1})$. $\square$

**Corollary 3.2.2.** $Fr : \overline{U}_1 \to Z_\varepsilon$ *is an isomorphism of Hopf Poisson algebras.*

In general, Lemma 3.2.3 implies

**Lemma 3.2.4.** $Fr(1 + \sum_{k=1}^\infty \overline{\psi}_k^{(i)} t^k) = \exp(\ell(\varepsilon^{d_i} - \varepsilon^{-d_i}) \sum_{k=1}^\infty E_{k\ell}^{(i)} t^k)$ *for* $i = 1 \ldots n$. $\square$

3.3 Introduce the notation:

$$X_i^- = -\sum_{k=-\infty}^{+\infty} \overline{E}_{-\alpha_i + k\delta} t^k, \quad X_i^+ = \sum_{k=-\infty}^{+\infty} \overline{E}_{\alpha_i + k\delta} t^k,$$

$$X_{>a,i}^\pm = \pm \sum_{k>a} \overline{E}_{\pm\alpha_i + k\delta} t^k, \quad X_{\leq a,i}^\pm = \pm \sum_{k \leq a} \overline{E}_{\pm\alpha_i + k\delta} t^k, \ \text{etc.},$$

$$\Psi_i = 1 + \sum_{k=1}^\infty \overline{\psi}_k^{(i)} t^k, \quad \Phi_i = 1 - \sum_{k=1}^\infty \overline{\psi}_{-k}^{(i)} t^{-k}, \ i = 1, \ldots, n.$$

We record the following calculations using the Poisson structure on $\overline{U}_1$ (see [Be] 4.1, 4.7 for the relevant commutation formulas) :

(3.3.1)
$$\{\overline{E}_j, \Phi_i\} = (\alpha_j|\alpha_i)\Phi_i X^+_{<0,j}, \qquad \{\overline{E}_j, \Psi_i\} = -(\alpha_j|\alpha_i)\Psi_i X^+_{>0,j},$$
$$\{\overline{F}_j, \Phi_i\} = -(\alpha_j|\alpha_i)\Phi_i X^-_{<0,j}, \qquad \{\overline{F}_j, \Psi_i\} = (\alpha_j|\alpha_i)\Psi_i X^-_{>0,j},$$
$$\{\overline{E}_i, K_i\} = -d_i \overline{E}_i K_i, \qquad \{\overline{F}_i, K_i\} = d_i \overline{F}_i K_i,$$
$$\{\overline{E}_i, X^+_{\geq 0,i}\} = -d_i X^+_{\geq 0,i} X^+_{>0,i}, \qquad \{\overline{E}_i, X^+_{\leq 0,i}\} = d_i X^+_{\leq 0,i} X^+_{<0,i},$$
$$\{\overline{E}_i, X^-_{>0,i}\} = d_i(K_i(\Psi - 1)), \qquad \{\overline{E}_i, X^-_{\leq 0,i}\} = d_i(K_i - K_i^{-1}\Phi_i),$$
$$\{\overline{F}_i, X^-_{\leq 0,i}\} = -d_i X^-_{\leq 0,i} X^-_{<0,i}, \qquad \{\overline{F}_i, X^-_{>0,i}\} = d_i X^-_{\geq 0,i} X^-_{>0,i},$$
$$\{\overline{F}_i, X^+_{\geq 0,i}\} = d_i(K_i^{-1} - K_i \Psi_i), \qquad \{\overline{F}_i, X^+_{<0,i}\} = d_i(K_i^{-1}\Phi_i - K_i^{-1}),$$
$$\{\overline{F}_i, X^+_i\} = d_i(K_i^{-1}\Phi_i - K_i \Psi_i).$$

Since $\Phi$ and $\Psi$ are series that start with 1, fractional powers of these series are well defined. Let $(\overline{a}_{ij}) = ({}^t A)^{-1}$, then $\omega_i = \sum_j \overline{a}_{ij}\alpha_j$. Define $\Phi_{\omega_i} = \prod_j \Phi_j^{\overline{a}_{ij}}$ (resp. $\Psi_{\omega_i} = \prod_j \Psi_j^{\overline{a}_{ij}}$). The following identities hold:

(3.3.2)
$$\{\overline{E}_j, \Phi_{\omega_i}\} = (\omega_i|\alpha_j)\Phi_{\omega_i} X^+_{<0,j}, \qquad \{\overline{E}_j, \Psi_{\omega_i}\} = -(\omega_i|\alpha_j)\Psi_{\omega_i} X^+_{>0,j},$$
$$\{\overline{F}_j, \Phi_{\omega_i}\} = -(\omega_i|\alpha_j)\Phi_{\omega_i} X^-_{<0,j}, \qquad \{\overline{F}_j, \Psi_{\omega_i}\} = (\omega_i|\alpha_j)\Psi_{\omega_i} X^-_{>0,j}.$$

## §4. The Poisson proalgebraic group $\Omega$.

4.1 Consider the group (cf. 1.3):
$$\tilde{\tilde{G}} = \underline{G}((t^{-1})) \times \underline{G}((t))$$

and introduce subgroups $\Omega$ and $K$ of $\tilde{\tilde{G}}$ as follows.

Let $\tilde{N}_- = \{g(t^{-1}) \in \underline{G}[[t^{-1}]] \mid g(\infty) \in N_-\}$, $\tilde{N}_+ = \{g(t) \in \underline{G}[[t]] \mid g(0) \in N_+\}$. Let $\Omega = \{(hu_-, h^{-1}u_+) \mid u_\pm \in \tilde{N}_\pm, h \in H\}$, $K = \{(g,g) \mid g \in \underline{G}[t, t^{-1}]\}$.

The Lie algebra of $\tilde{\tilde{G}}$ is $\tilde{\tilde{\mathfrak{g}}} := \mathfrak{g}((t^{-1})) \oplus \mathfrak{g}((t))$. The Lie subalgebra Lie $\Omega \subset \tilde{\tilde{\mathfrak{g}}}$ of $\Omega$ consists of pairs $(a_1(t^{-1}), a_2(t)) \in \tilde{\tilde{\mathfrak{g}}}$, where $a_1(t^{-1}) \in \mathfrak{g}[[t^{-1}]], a_2(t) \in \mathfrak{g}[[t]]$ are such that $a_1(\infty) = n_- + h, a_2(0) = n_+ - h$ and $n_\pm \in \mathfrak{n}_\pm, h \in \mathfrak{h}$. The Lie algebra Lie $K$ consists of pairs $(a, a)$, where $a \in \mathfrak{g}[t, t^{-1}]$. We have

(4.1.1)
$$\tilde{\tilde{\mathfrak{g}}} = \text{Lie } \Omega \oplus \text{Lie } K,$$

where $\oplus$ is the direct sum of vector spaces. The invariant bilinear symmetric form $(.|.)$ on the Lie algebra $\mathfrak{g}$ extend bilinearly to a $\mathbb{C}((t^{-1}))$(resp. $\mathbb{C}((t))$)–valued form on $\mathfrak{g}((t^{-1}))$ (resp. $\mathfrak{g}((t))$). We denote by $(.|.)_\infty$ (resp. $(.|.)_0$) the constant term. This is a $\mathbb{C}$–valued invariant bilinear symmetric form on $\mathfrak{g}((t^{-1}))$ (resp. $\mathfrak{g}((t))$). Define a ($\mathbb{C}$–valued) invariant bilinear symmetric form $(.|.)$ on $\tilde{\tilde{\mathfrak{g}}}$ by:

(4.1.2)
$$((x_1, x_2)|(y_1, y_2)) = -(x_1|y_1)_\infty + (x_2|y_2)_0.$$

then the subalgebras Lie $\Omega$ and Lie $K$ of $\tilde{\tilde{\mathfrak{g}}}$ are isotropic with respect to the form (4.1.2). Thus $(\tilde{\tilde{\mathfrak{g}}}, \text{Lie } \Omega, \text{Lie } K)$ is a Manin triple. This endows the proalgebraic group $\Omega$ with a canonical structure of a Poisson proalgebraic group (see e.g. [DC–K–P3, §4]).

The general description of symplectic leaves (given e.g. by [DC–K–P3, Proposition 4.2]) implies the following:

**Proposition 4.1.** *Consider the following action of the group $K \times K$ on $\tilde{\tilde{G}}$:*

(4.1.3) $$((a,a),(b,b)) \cdot (g_1, g_2) = (ag_1 b^{-1}, ag_2 b^{-1}).$$

*Then the symplectic leaves of $\Omega$ are connected components of intersections of orbits of this action with $\Omega \subset \tilde{\tilde{G}}$.* □

Note that the restriction of the canonical map

$$\alpha: \Omega \to \tilde{\tilde{G}}/K$$

is a finite covering of some open set of $\tilde{\tilde{G}}/K$. Considering the left action of $K$ on $\tilde{\tilde{G}}/K$, an element $e \in \mathrm{Lie}\, K$ defines a vector field on $\tilde{\tilde{G}}/K$ which using $\alpha$ can be lifted to $\Omega$. We shall denote again by $e$ the resulting vector field on $\Omega$. All these vector fields are tangent to the symplectic leaves of $\Omega$.

4.2 For each $i = 0, \ldots, n$ there exists a unique normal subgroup $\tilde{N}_+^{(i)}$ (resp. $\tilde{N}_-^{(i)}$) of $\tilde{N}_+$ (resp. $\tilde{N}_-$) such that $\tilde{N}_+ = \tilde{N}_+^{(i)} \ltimes \exp(\mathbb{C}e_i)$, $\tilde{N}_- = \tilde{N}_-^{(i)} \ltimes \exp(\mathbb{C}f_i)$. This allows us to define regular functions $\tilde{x}_i$ and $\tilde{y}_i$ on $\tilde{N}_+$ and $\tilde{N}_-$ respectively by letting:

$$u_+ = u_+^{(i)} \exp(-\tilde{x}_i e_i), \quad u_- = u_-^{(i)} \exp(\tilde{y}_i f_i), \text{ where } u_\pm^{(i)} \in \tilde{N}_\pm^{(i)}.$$

Since $\mathbb{C}[H] = P$, any $\alpha \in P$ defines a regular function on $H$, which we denote by $\tilde{z}_\alpha$. We extend these functions to regular functions on $\Omega$ by letting $\tilde{x}_i$, $\tilde{y}_i$ and $\tilde{z}_\alpha$ be defined at the point $(h^{-1}u_-, hu_+)$ by $\tilde{x}_i(u_+)$, $\tilde{y}_i(u_-)$, and $\alpha(h)$ respectively.

As in [DC–K–P] the braid group acts on $\Omega$ by the formula

$$\tilde{T}_i(hu_-, h^{-1}u_+) = (t_i h u_-^{(i)} \exp(\tilde{x}_i e_i) t_i^{-1}, t_i h (\exp \tilde{y}_i f_i) h^{-2} u_+^{(i)} t_i^{-1}).$$

For an element $a$ of a Poisson algebra $A$ we denote by $P_a$ the derivation of $A$ defined by $P_a(x) = \{a, x\}$. In the same way as in [DC–K–P2, § 7.6] the following theorem is proved.

**Theorem 4.2** (a) *The functions $\tilde{x}_i$, $\tilde{y}_i$, $(i = 0, \ldots n)$ and $\tilde{z}_\alpha$ ($\alpha \in P$) generate the coordinate ring $\mathbb{C}[\Omega]$ as a Poisson algebra.*
  (b) $\Delta \tilde{x}_i = 1 \otimes \tilde{x}_i + \tilde{x}_i \otimes \tilde{z}_{-\alpha_i}$, $\Delta \tilde{y}_i = 1 \otimes \tilde{y}_i + \tilde{y}_i \otimes \tilde{z}_{-\alpha_i}$, $\Delta \tilde{z}_\alpha = \tilde{z}_\alpha \otimes \tilde{z}_\alpha$.
  (c) $P_{\tilde{z}_{\alpha_i}} \tilde{x}_i = -d_i \tilde{z}_{\alpha_i}(f_i, f_i)$, $P_{\tilde{z}_{\alpha_i}} \tilde{y}_i = d_i \tilde{z}_{\alpha_i}(e_i, e_i)$, $P_{\tilde{z}_{\alpha_i}} = \frac{1}{2} d_i \tilde{z}_{\alpha_i}(\alpha_i^\vee, \alpha_i^\vee)$.
  (d) *The $\tilde{T}_i$ define a map $\tilde{\mathcal{B}} \to \mathrm{Aut}(\Omega)$ (where Aut denotes Poisson algebraic variety automorphisms).* □

## §5. The isomorphism $\pi: \mathrm{Spec}\, Z_\varepsilon \to \Omega$.

5.1 Consider the following (closed proalgebraic) subgroups of the group $\tilde{N}_+$:

$$\tilde{N}_+^+ = \prod_{k \leq 0} \exp(\mathbb{C}e_{\beta_k}), \; \tilde{N}_+^- = \prod_{k > 0} \exp(\mathbb{C}e_{\beta_k}), \; \tilde{N}_+^0 = \prod_{i=1}^n \prod_{k \geq 1} \exp(\mathbb{C}\omega_k^{\vee(i)}),$$

where we let $\omega_k^{\vee(i)} = t^k \otimes \omega_i^\vee$, and similarly those of $\tilde{N}_-$:

$$\tilde{N}_-^+ = \prod_{k > 0} \exp(\mathbb{C}e_{-\beta_k}), \; \tilde{N}_-^- = \prod_{k \leq 0} \exp(\mathbb{C}e_{-\beta_k}), \; \tilde{N}_+^0 = \prod_{i=1}^n \prod_{k \geq 1} \exp(\mathbb{C}\omega_{-k}^{\vee(i)}).$$

Then multiplication establishes isomorphisms of proalgebraic varieties:

(5.1.1) $$\tilde{N}_\pm \simeq \tilde{N}_\pm^- \times \tilde{N}_\pm^0 \times \tilde{N}_\pm^+.$$

Define functions $\tilde{x}_{\beta_k}$ and $\tilde{x}_k^{(i)}$ by letting

$$\tilde{x}_{\beta_k}(\prod_j \exp \gamma_j e_{\beta_j}) = \gamma_k, \ \tilde{x}_k^{(i)}(\prod_j \exp \gamma_j^{(i)} \omega_j^{\vee(i)}) = \gamma_k^{(i)}.$$

Then the coordinate ring of the group $\tilde{N}_+^{+(\text{resp. } -)}$ is the polynomial algebra $\mathbb{C}[\tilde{x}_{\beta_k} \mid k > 0$ (resp. $k \leq 0)]$, and similar statements holds for the groups $\tilde{N}_-^\pm$. Finally, the coordinate ring of $\tilde{N}_\pm^0$ is the algebra $\mathbb{C}[\tilde{x}_{\pm k}^{(i)}, (k > 0)]$.

5.2 For $\alpha \in \Delta$ and $i = 1, \ldots, n$ introduce constants:

$$d_\alpha = \frac{1}{2}(\alpha|\alpha), \ c_\alpha = (\varepsilon^{d_\alpha} - \varepsilon^{-d_\alpha})^\ell, \ b_i = \ell(\varepsilon^{d_i} - \varepsilon^{-d_i}).$$

Consider the following elements of $Z_\varepsilon$:

$$z_\beta = K_\beta^\ell \ (\beta \in P); \ x_{-\alpha+k\delta} = c_\alpha E_{-\alpha+k\delta}^\ell \ (\alpha \in \Delta_+, \ k \in \mathbb{Z}),$$

$$x_{\alpha+k\delta} = -c_\alpha z_{-\alpha} E_{\alpha+k\delta}^\ell \ (\alpha \in \Delta_+, \ k \in \mathbb{Z}); x_{k\delta}^{(i)} = b_i E_{\ell k\delta}^{(i)} \ (k \in \mathbb{Z}^\times).$$

Introduce the subalgebras $Z_0^0, Z_+^+, Z_+^0$ and $Z_+^-$ of $Z_\varepsilon$ generated by the elements $z_\beta$ ($\beta \in P$); $x_{\beta_k}$ ($k > 0$); $x_{-\beta_k}$ ($k \geq 0$); and $x_{k\delta}^{(i)}$ ($i = 1, \ldots, n, \ k \in \mathbb{Z}^\times$). Similarly introduce the subalgebras $Z_-^+, Z_-^0$, and $Z_-^-$ of $Z_\varepsilon$. By Theorem 5.2, $Z_\varepsilon$ is isomorphic to the tensor product of these subalgebras. Hence defining the subalgebras $Z_+ = Z_+^- \otimes Z_+^0 \otimes Z_+^+$, and $Z_- = Z_-^- \otimes Z_-^0 \otimes Z_-^+$ we have:

(5.2.1) $$Z_\varepsilon = Z_- \otimes Z_0^0 \otimes Z_+.$$

As usual, given a commutative associative algebra $A$ over $\mathbb{C}$ we denote by Spec $A$ the proalgebraic variety of all algebra homomorphisms $A \to \mathbb{C}$. Note that given a proalgebraic variety $X$, defining a regular map $\rho : \text{Spec } A \to X$ amounts to giving an element of $X(A)$, which we denote by the same letter $\rho$.

We let

(5.2.2)
$$\pi_+^+ = \prod_{k \leq 0, <} \exp(-x_{\beta_k} e_{\beta_k}) \in \tilde{N}_+^+(Z_+^+), \ \pi_-^+ = \prod_{k > 0, <} \exp(z_{\beta_k}^2 x_{-\beta_k} e_{-\beta_k}) \in \tilde{N}_-^+(Z_-^+),$$

$$\pi_+^- = \prod_{k > 0, <} \exp(-x_{\beta_k} e_{\beta_k}) \in \tilde{N}_+^-(Z_+^-), \ \pi_-^- = \prod_{k \leq 0, <} \exp(x_{-\beta_k} e_{-\beta_k}) \in \tilde{N}_-^-(Z_-^-),$$

$$\pi_+^0 = \prod_{i=1}^n \prod_{k > 0} \exp(x_{k\delta}^{(i)} t^k \omega_i^\vee) \in \tilde{N}_+^0(Z_+^0), \ \pi_-^0 = \prod_{i=1}^n \prod_{k > 0} \exp(-x_{-k\delta}^{(i)} t^{-k} \omega_i^\vee) \in \tilde{N}_-^0(Z_-^0).$$

As previously remarked, (5.2.2) defines maps $\pi_+^+ : \text{Spec } Z_+^+ \to \tilde{N}_+^+$, etc. Finally we define the map $\pi_0^0 : \text{Spec } Z_0^0 \to H$ by identifying the function $\alpha$ on $H$ with the element $z_\alpha \in Z_0^0$. We may write $\pi_0^0$ in the form

(5.2.3) $$\pi_0^0 = \sum_i z_{\omega_i} \otimes \alpha_i^\vee = \sum_i z_{\alpha_i} \otimes \omega_i^\vee \in Z_0^0 \otimes_\mathbb{Z} Q^\vee = \underline{H}(Z_0^0).$$

We point out that $\pi_+^0$ (resp. $\pi_-^0$) is the image under the Frobenius correspondence of the map $\sum_i \Psi_{\omega_i} \otimes \alpha_i^\vee$ (resp. $\sum_i \Phi_{\omega_i} \otimes \alpha_i^\vee$). This is since

$$\sum_i \mathrm{Fr}(\Psi_{\omega_i}) \otimes \alpha_i^\vee = \prod_{i=1}^n \prod_{k \geq 0} \prod_{j=1}^n \exp(x_{k\delta}^{(j)} t^k \alpha_i^\vee)^{\overline{a}_{ij}} = \prod_{i,j} \prod_k \exp(\overline{a}_{ij} x_{k\delta}^{(j)} \alpha_i^\vee)$$
$$= \prod_{i,j} \prod_k \exp(x_{k\delta}^{(j)} \overline{a}_{ij} \alpha_i^\vee) = \prod_j \prod_k \exp(x_{k\delta}^{(j)} \omega_j^\vee).$$

We define the maps:

(5.2.4)
$$\pi_+ = \pi_+^+ \times \pi_+^0 \times \pi_+^- : \mathrm{Spec}\, Z_+ \to \tilde{N}_+,$$
$$\pi_- = \pi_-^+ \times \pi_-^0 \times \pi_-^- : \mathrm{Spec}\, Z_- \to \tilde{N}_-.$$

Using (5.2.1), we write an element of $\mathrm{Spec}\, Z_\varepsilon$ in the form $u_- h u_+$, where $u_\pm \in \mathrm{Spec}\, Z_\pm, h \in \mathrm{Spec}\, Z_0^0$. Now we may define the isomorphism of proalgebraic varieties

$$\pi : \mathrm{Spec}\, Z_\varepsilon \xrightarrow{\sim} \Omega$$

by letting

(5.2.5)
$$\pi(u_- h u_+) = \big((\pi_0^0(h))^{-1} \pi_-(u_-), \pi_0^0(h) \pi_+(u_+)\big).$$

**Remark 5.2.** We have written $\pi$ in the form (5.2.5) so that its relationship to the finite type map [DC–K–P] is apparent. In later consideration of finite dimensional representations it will be useful to express $\pi$ in a form where $(\pi_0^0)^{\pm 1}$ is incorporated into $\pi_-$ and $\pi_+$ in the first and second factors respectively.

5.3 The following is our first key result.

**Theorem 5.3.** *The map $\pi$ is an isomorphism of Poisson proalgebraic groups which commutes with the action of $\tilde{\mathcal{B}}$.*

Proof of this theorem is along the same lines as the analogous result for the finite type case in [DC–K–P]. It is based on the same simple lemma:

**Lemma 5.3.** *[DC–K–P2, Lemma 7.2] Let A and B be two commutative Poisson Hopf algebras and let $\varphi : A \to B$ be an algebra isomorphism compatible with the augmentation maps. Suppose that elements $a_1, \ldots, a_s$ generate A as a Poisson algebra and that the following two properties hold:*

*(i) $(\varphi \otimes \varphi)\Delta_A(a_i) = \Delta_B \varphi(a_i)$, $i = 1, \ldots, s$;*
*(ii) $\{\varphi(a_i), \varphi(a)\} = \varphi(\{a_i, a\})$, $i = 1, \ldots, s, a \in A$.*

*Then $\varphi$ is an isomorphism of Poisson Hopf algebras.* □

We apply this lemma to Poisson Hopf algebras $A = \mathbb{C}[\Omega]$, $B = Z_\varepsilon$, and the map $\varphi = \pi^*$. For the Poisson generators of $A$ we take elements $\tilde{x}_i, \tilde{y}_i$, $(i = 0, \ldots, n)$ and $\tilde{z}_\alpha$ $(\alpha \in P)$ (cf. Theorem 4.2(a)). Note that $\varphi(\tilde{x}_i) = x_i$, $\varphi(\tilde{y}_i) = y_i$, and $\varphi(\tilde{z}_\alpha) = z_\alpha$ $(\alpha \in P)$, that $\varphi$ is compatible with augmentation maps and that the assumption $(i)$ of Lemma 5.3 obviously holds (cf. Theorem 4.2(b)). Hence, in view of Theorem 4.2(c), in order to check assumption (ii) of Lemma 5.3 we have to show that for $i = 0, \ldots, n$:

$$(5.3.1) \qquad P_{z_{\alpha_i} x_i} = -d_i z_{\alpha_i}(f_i, f_i), \ P_{z_{\alpha_i} y_i} = d_i z_{\alpha_i}(e_i, e_i), \ P_{z_{\alpha_i}} = \frac{1}{2} d_i z_{\alpha_i}(\alpha_i^\vee, \alpha_i^\vee),$$

where the vector fields $(f_i, f_i)$, etc. on $\operatorname{Spec} Z_\varepsilon$ are the pull–backs of the corresponding vector fields on $\Omega$ via the map $\varphi$.

In order to prove (5.3.1) consider the map

$$\gamma((a, b)) = a^{-1} b, \ (a, b) \in \Omega.$$

Then the action (4.1.3) of $K$ on $\widetilde{\widetilde{G}}$ induces the action by conjugation of $\underline{G}[t, t^{-1}]$ on $\gamma(\Omega)$. Since the fibers of $\gamma$ are finite, it suffices to check the pushdowns of the equalities (5.3.1) to $\gamma(\Omega)$. It is easy to see that the latter equalities are as follows $(i = 0, \ldots, n)$:

$$(5.3.2) \qquad P_{z_{\alpha_i} x_i} = -d_i z_{\alpha_i} f_i, \ P_{z_{\alpha_i} y_i} = d_i z_{\alpha_i} e_i, \ P_{z_{\alpha_i}} = \frac{1}{2} d_i z_{\alpha_i} \alpha_i^\vee.$$

Consider a faithful finite–dimensional representation of the group $G$. Then the meaning of, for example, the first equality of (5.3.2) is interpreted as follows. The left–hand side is the Poisson bracket of $z_{\alpha_i} x_i$ with all elements of the matrix

$$(5.3.3) \qquad M := (\pi_-^- \pi_-^0 \pi_-^+)^{-1} (\pi_0^0)^2 (\pi_+^- \pi_+^0 \pi_+^+).$$

The right–hand side is the usual bracket $[-d_i z_{\alpha_i} f_i, M]$ (since $\underline{G}[t, t^{-1}]$ acts by conjugation).

We now explain how to perform these calculations assuming the affine rank 2 case, which will be calculated in the next section making use of the Frobenius map. Recall that (5.3.2) holds for the finite type case (this is the main result of [DC–K–P]).

Let $Z_\varepsilon^{0,\mathrm{im}}$ be the subalgebra of $Z_\varepsilon$ generated by $z_\beta$ and $x_{k\delta}^{(i)}$ ($\beta \in Q^\vee$, $k \in \mathbb{Z}^\times, i = 1, \ldots, n$). Recall that for each $\alpha \in \widetilde{\Delta}_+^{\mathrm{re}}$, $x_\alpha$ is defined by applying some braid group operators corresponding to an initial of the reduced expression $\prod_{k<0} s_{i_k}$, or $\prod_{k \geq 0} s_{i_k}$ (see §4). By the same proof as [DC–K–P, Proposition 2] we have:

**Lemma 5.3.1.** *Let $x'_\alpha$, $x'_{-\alpha}$ ($\alpha \in \widetilde{\Delta}_+$) be defined as in (1.6.1) and §5.2 where $\prod_{k \geq 0} s_{i_k}$ (resp. $\prod_{k<0} s_{i_k}$) (see (1.6.1)) is replaced by a new expression obtained by substituting an arbitrary set of braid relations. Then $x'_\alpha, x'_{-\alpha}$ generate $Z_\varepsilon$ over $Z_\varepsilon^{0,\mathrm{im}}$.* □

**Lemma 5.3.2.** *The derivations $P_{z_{\alpha_i} x_i}$ and $-d_i z_{\alpha_i} f_i$ coincide on $x_{k\delta}^{(j)}$ ($j = 1, \ldots, n$).*

*Proof.* We reduce this to the rank two calculation as follows. Consider the map $M_i$ defined by modifying $\pi_-^-, \pi_-^+, \pi_+^-$ and $\pi_+^+$ in (5.3.3) to contain only factors corresponding to root vectors of the form $\pm \alpha_i \pm k\delta$. Then $M_i = \overline{M} B_i$ where $\overline{M}$ is as in (5.3.3) for the case of $\widetilde{\mathfrak{sl}}_2$ and

$$(5.3.4) \qquad B_i = \exp\bigl( \sum_{j \neq i, k > 0} x_{k\delta}^{(j)} t^k \omega_j^\vee \bigr).$$

It follows from (5.3.1) that both $P_{z_{\alpha_i} x_i}$ and $-d_i z_{\alpha_i} f_i$ act as 0 on $B_i$, and therefore they coincide on $M_i$ if they coincide on $\overline{M}$.

In order to prove the lemma we must show that $P_{z_{\alpha_i} x_i}$ and $-d_i z_{\alpha_i} f_i$ coincide on Spec $Z_\varepsilon$ when we consider $-d_i z_{\alpha_i} f_i$ when pulled back via the map $M$. Consider the subvariety $\Omega^{(i)}$ of $\Omega$ defined by replacing $G$ in 4.1 by the connected subgroup of $G$ whose Lie algebra is $\mathfrak{h} \oplus \mathbb{C} e_i \oplus \mathbb{C} f_i$. Then $\Omega^{(i)}$ is characterized as the subvariety of $\Omega$ for which $x_\alpha(p) = 0$ when $\alpha \neq \pm \alpha_i + k\delta$ ($p \in \Omega$, $k \in \mathbb{Z}$, $\alpha \in \widetilde{\Delta}^{\text{re}}$). Let Spec $Z_\varepsilon^{(i)}$ be the Poisson subvariety of Spec $Z_\varepsilon$ consisting of $p \in$ Spec $Z_\varepsilon$ for which $x_\alpha(p) = 0$ when $\alpha \neq \pm \alpha_i + k\delta$ ($p \in \Omega$, $k \in \mathbb{Z}$, $\alpha \in \widetilde{\Delta}^{\text{re}}$). Then $\gamma \circ \pi_{|\text{Spec } Z_\varepsilon^{(i)}} = M_i$. Using Frobenius it is clear that $Z_\varepsilon^{(i)}$ is a Poisson subalgebra of $Z_\varepsilon$. Since $\Omega^{(i)}$ and Spec $Z_\varepsilon^{(i)}$ are respectively invariant when integrating along $f_i$ and $P_{z_{\alpha_i} x_i}$ the lemma follows. $\square$

Since the derivations $P_{z_{\alpha_i} x_i}$ and $-d_i z_{\alpha_i} f_i$ coincide on $Z_\varepsilon^{0,\text{im}}$ by the rank 2 calculations, it suffices to show that

(5.3.5) $$P_{z_{\alpha_i} x_i}(x_\alpha) = -d_i z_{\alpha_i} f_i(x_\alpha),\ \alpha \in \widetilde{\Delta}^{\text{re}},\ i = 1, \ldots, n.$$

We give a proof assuming that $\alpha = \bar{\alpha} + k\delta$ where $k \geq 0$ and $\bar{\alpha} \in \Delta_+$. The cases where $\alpha \in \widetilde{\Delta}_-^{\text{re}}$ or $\bar{\alpha} \in \Delta_-$ are similar. Given two non–proportional roots $\alpha$ and $\beta$, we denote by $R_{\alpha,\beta}$ the intersection of the $\mathbb{Z}$–span of $\alpha$ and $\beta$ with $\widetilde{\Delta}^{\text{re}}$ and let $R_{\alpha,\beta}^+ = R_{\alpha,\beta} \cap \widetilde{\Delta}_+^{\text{re}}$. Then $R_{\alpha,\beta}$ is a rank 2 root system with $R_{\alpha,\beta}^+$ being a subset of positive roots.

We consider the following two possibilities for $\alpha$:

(a) $\alpha = \alpha_i + k\delta$, or
(b) $\alpha$ and $\alpha_i$ generate a subroot system of finite type.

In case (a) the statement follows from the $U_q(\widetilde{\mathfrak{sl}_2})$ calculations in the next section and the Drinfel'd relations (1.4.5) which show these calculations hold for all $i = 1, \ldots, n$. Assume case (b) holds. There exist two simple roots, $\alpha_1, \alpha_2 \in \Pi$ and there exists $y \in Q^\vee$, $w \in W_0$ such that

$$yw R_{\alpha_1, \alpha_2}^+ = R_{\alpha_i, \alpha}^+ \text{ and } yw(\alpha_1) = \alpha_i.$$

Fix a reduced expression of $w = s_{i_1} \ldots s_{i_k}$. Let $w_0' = s_1 s_2 s_1 \ldots s_\varepsilon$, where $\varepsilon = 1$ or 2, be the reduced expression $J'$ of the longest element of the Weyl group of $R_{\alpha_1,\alpha_2}$ and let $m = \ell(w_0')$. Then the expression $ww_0' = s_{i_1} \ldots s_{i_k} s_1 s_2 \ldots s_\varepsilon$ is reduced. By [Pa, Proposition 7] it is possible to complete the reduced expression $yww_0$ to a reduced expression of some positive power of $x^{-1}$ (where $x$ is as in 1.4). Then $\widetilde{\Delta}_+^{\text{re}}$ breaks into five pieces:

(5.3.6)
$$R^1 := \{\beta_0, \ldots, \beta_{-k}\},\ \beta_{-k-1} = \alpha_i,$$
$$R^2 := \{\beta_{-k-2}, \ldots, \beta_{-k-m}\} = R_{\alpha_i,\alpha}^+,\ R^3 := \{\beta_{-k-m-1}, \ldots, k\delta, \ldots, \beta_l, l > 0\}.$$

Let $\mathfrak{g}_\pm^i = \mathfrak{h} \otimes \mathbb{Z}[t, t^{-1}] \bigoplus_{\gamma \in R^i} \mathbb{C} e_{\pm \gamma}$, $i = 1, 2, 3$. These are subalgebras of the Lie algebra $\mathfrak{g}$ normalized by the 3–dimensional subalgebra $\mathbb{C} e_i + \mathbb{C} h_i + \mathbb{C} f_i$. This is so because it follows from [Pa, Theorem 1] that $R^i \pm \alpha_i \subset R^i$, for $i = 1, 2, 3$. Let $U_\pm^i$ be the subgroups of $U_\pm$ corresponding to the $\mathfrak{g}_\pm^i$.

We turn now to the map $M$ which we decompose according to the above decomposition of $\widetilde{\Delta}_+^{\text{re}}$:

$$M = \pi_-^+ \times \pi_-^0 \pi_-^3 (\exp x_{-\alpha_i})) \pi_-^1 \times (\pi_0^0)^2 \pi_+^1 (\exp x_{\alpha_i} e_i) \pi_+^2 \pi_+^3 \pi_+^0 \pi_+^-$$
$$= \pi_-^+ \times \pi_-^0 \pi_-^3 \pi_-^{1'} (\exp x_{-\alpha_i})) \times (\pi_0^0)^2 (\exp x_{\alpha_i} e_i) \pi_+^{1'} \pi_+^2 \pi_+^3 \pi_+^0 \pi_+^-,$$

where $\pi_-^{1\prime} = (\exp x_{-\alpha_i} e_{-\alpha_i}) \pi_-^1 (\exp -x_{-\alpha_i} e_{-\alpha_i}) \subset U_-^1$ and $\pi_+^{1\prime} = (\exp x_i e_i) \pi_+^1 (\exp -x_i e_i) \in U_+^1$.

Consider the subalgebra $Z_0^{1,2}$ of $Z_\varepsilon$ generated over $Z_0^0$ by all $x_\gamma$ and $x_{-\gamma}$ with $\gamma \in R_{\alpha_1,\alpha_2}^+$. We want to prove the following formula:

$$(5.3.7) \qquad (d_i z_i f_i)(T_w(a)) = T_w(d_1 z_1 f_1(a)) \text{ for } a \in Z_0^{1,2}.$$

This formula implies (5.3.5) using the calculations in the rank 2 case, we have for $a \in Z_0^{1,2}$: $P_{z_{\alpha_i} x_i}(T_w(a)) = T_w P_{z_{\alpha_1} x_1}(a) = T_w(z_1 f_1(a))$.

In order to prove (5.3.7) note that the action of $z_i f_i$ on $Z_\varepsilon$ may be calculated as follows. Write for $t \in \mathbb{C}$:

$$(\exp t z_i f_i) \pi (\exp -t z_i f_i) = \prod_k (\exp x_{-\beta_k}(t) e_{-\beta_k}) \prod_k (\exp x_{\beta_k}(t) e_{\beta_k}).$$

Then $f_i(x_{\beta_k}) = \frac{d}{dt} x_{\beta_k}(t)|_{t=0} \, \beta \in \widetilde{\Delta}_+^{\text{re}}$, and similarly for $x_{-\beta_k}$.

But $x_\alpha$ (resp. $x_{-\alpha}$) occurs only in $\pi_2^+$ (resp. $\pi_2^-$) and all other factors of $\pi$ lie in the subgroups normalized by $\exp t z_i f_i$ and having trivial intersection with $U_+^2$ (resp. $U_-^2$). Thus, it suffices to perform the calculation in $U_+^2$ (resp. $U_-^2$). We have:

$$\prod_{s=k+2}^{k+m} \exp x_\beta^J(t) e_\beta^J = (\exp t z_i f_i) \prod_{s=2}^m \exp T_w(x_s^{J'} e_s^{J'})(\exp -t z_i f_i)$$

$$= T_w((\exp t z_1 f_1)(\prod_{s=2}^m \exp x_s^{J'} e_s^{J'})(\exp -t z_1 f_1)),$$

and we can use the calculation in the finite type rank 2 case in [DC–K–P]. □

5.4 In this section we make explicit the calculations for $U_q(\widetilde{\mathfrak{sl}}_2)$, and $U_q^i(\widetilde{\mathfrak{sl}}_2)$ which will imply Theorem 5.3.

In the case of $\widetilde{\mathfrak{sl}}_2$ the convex order (1.1.1) is as follows:

$$(5.4.1) \qquad \delta - \alpha < 2\delta - \alpha < \cdots < 2\delta < \delta < \cdots < \alpha + \delta < \alpha.$$

Let $h_\pm(t) = \exp(\frac{1}{2} \sum_{k=1}^\infty x_{\pm k\delta} t^{\pm k})$, where $x_{k\delta} = x_{k\delta}^{(1)}$. Then the map (5.2.10) can be written as follows:

$$(5.4.2) \qquad \left( \begin{pmatrix} z_\omega^{-1} & 0 \\ 0 & z_\omega \end{pmatrix} \begin{pmatrix} 1 & \sum_{k=1}^\infty z_\alpha^2 x_{-k\delta+\alpha} t^{-k} \\ 0 & 1 \end{pmatrix} \begin{pmatrix} h_-(t)^{-1} & 0 \\ 0 & h_-(t) \end{pmatrix} \begin{pmatrix} 1 & 0 \\ \sum_{k=0}^\infty x_{-\alpha-k\delta} t^{-k} & 1 \end{pmatrix} \right.,$$

$$\left. \begin{pmatrix} z_\omega & 0 \\ 0 & z_\omega^{-1} \end{pmatrix} \begin{pmatrix} 1 & \sum_{k=0}^\infty -x_{\alpha+k\delta} t^k \\ 0 & 1 \end{pmatrix} \begin{pmatrix} h_+(t) & 0 \\ 0 & h_+(t)^{-1} \end{pmatrix} \begin{pmatrix} 1 & 0 \\ \sum_{k=1}^\infty -x_{-\alpha+k\delta} t^k & 1 \end{pmatrix} \right).$$

Pulling this map back to $\overline{U}_1$ via the Frobenius map we can consider the corresponding map to (5.1.2):

$$\left( \begin{pmatrix} K_\omega^{-1} & 0 \\ 0 & K_\omega \end{pmatrix} \begin{pmatrix} 1 & -K_\alpha X_{\leq -1}^+ \\ 0 & 1 \end{pmatrix} \begin{pmatrix} \sqrt{\Phi} & 0 \\ 0 & \sqrt{\Phi}^{-1} \end{pmatrix} \begin{pmatrix} 1 & 0 \\ -X_{\leq 0}^- & 1 \end{pmatrix} \right.,$$

(5.4.3)
$$\begin{pmatrix} K_\omega & 0 \\ 0 & K_\omega^{-1} \end{pmatrix} \begin{pmatrix} 1 & K_\alpha^{-1} X_{\geq 0}^+ \\ 0 & 1 \end{pmatrix} \begin{pmatrix} \sqrt{\Psi} & 0 \\ 0 & \sqrt{\Psi}^{-1} \end{pmatrix} \begin{pmatrix} 1 & 0 \\ X_{>0}^- & 1 \end{pmatrix}.$$

Note that (5.4.3) can be rewritten as

(5.4.4)
$$\left( \begin{pmatrix} 1 & -X_{\leq -1}^+ \\ 0 & 1 \end{pmatrix} \begin{pmatrix} K_\omega^{-1} \sqrt{\Phi} & 0 \\ 0 & K_\omega \sqrt{\Phi}^{-1} \end{pmatrix} \begin{pmatrix} 1 & 0 \\ -X_{\leq 0}^- & 1 \end{pmatrix}, \right.$$
$$\left. \begin{pmatrix} 1 & X_{\geq 0}^+ \\ 0 & 1 \end{pmatrix} \begin{pmatrix} K_\omega \sqrt{\Psi} & 0 \\ 0 & K_\omega^{-1} \sqrt{\Psi}^{-1} \end{pmatrix} \begin{pmatrix} 1 & 0 \\ X_{>0}^- & 1 \end{pmatrix} \right)$$

(5.4.5)
$$= \left( \begin{pmatrix} K_\omega^{-1} \sqrt{\Phi} + K_\omega \sqrt{\Phi}^{-1} X_{\leq 0}^- X_{\leq -1}^+ & -K_\omega \sqrt{\Phi}^{-1} X_{\leq -1}^+ \\ -K_\omega \sqrt{\Phi}^{-1} X_{\leq 0}^- & K_\omega \sqrt{\Phi}^{-1} \end{pmatrix}, \right.$$
$$\left. \begin{pmatrix} K_\omega \sqrt{\Psi} + K_\omega^{-1} \sqrt{\Psi}^{-1} X_{\geq 0}^+ X_{>0}^- & K_\omega^{-1} \sqrt{\Psi}^{-1} X_{\geq 0}^+ \\ K_\omega^{-1} \sqrt{\Psi}^{-1} X_{>0}^- & K_\omega^{-1} \sqrt{\Psi}^{-1} \end{pmatrix} \right).$$

We consider the composition of map (5.3.3) with $Fr^{-1}$ where $\gamma(a,b) = a^{-1}b$. In the case of $\widetilde{\mathfrak{sl}}_2$ we have:

$$A = \begin{pmatrix} K_1 \sqrt{\Psi} \sqrt{\Phi}^{-1} + \sqrt{\Phi}^{-1} \sqrt{\Psi}^{-1} X^+ X_{>0}^- & \sqrt{\Psi}^{-1} \sqrt{\Phi}^{-1} X^+ \\ K_1 \sqrt{\Phi}^{-1} \sqrt{\Psi} X_{\leq 0}^- + K_1^{-1} \sqrt{\Phi} \sqrt{\Psi}^{-1} X_{>0}^- & K_1^{-1} \sqrt{\Phi} \sqrt{\Psi}^{-1} + \sqrt{\Phi}^{-1} \sqrt{\Psi}^{-1} X^+ X_{\leq 0}^- \\ + \sqrt{\Phi}^{-1} \sqrt{\Psi}^{-1} X_{\leq 0}^- X_{>0}^- X^+ & \end{pmatrix}$$

Then using the Frobenius map to reinterpret (5.3.2) for $i = 0, 1$, we have $P_{-\overline{E}_i} = -d_i K_i f_i$, $P_{\overline{F}_i K_i} = d_i K_i e_i$, where we let $K_0 = K_1^{-1}$.

**Proposition 5.4.**

a) $\{\overline{E}_1, A\} = -K_1[e_{21}, A]$,  $\{\overline{E}_0, A\} = -K_0[t^{-1}e_{12}, A]$,

b) $\{\overline{F}_1 K_1, A\} = K_1[e_{12}, A]$,  $\{\overline{F}_0 K_0, A\} = K_0[te_{21}, A]$,

c) $\{K_1, A\} = \frac{1}{2} K_1[e_{11} - e_{22}, A]$, $\{K_0, A\} = \frac{1}{2} K_0[e_{22} - e_{11}, A]$.

*Proof.* a), b) and c) follow from the formulas (3.3.1) by explicit calculation.  □

### §6 On the parametrization of finite–dimensional irreducible representations.

6.1 We recall the following material which is useful for studying finite dimensional representations of $\widetilde{U}_\varepsilon$.

**Lemma 6.1 (cf. [S, ch IV]).** *Consider the polynomial $Q(x) = 1 + a_1 x + \ldots a_d x_d$ over $\mathbb{C}$ of degree $d$. Then for a sequence of complex $N \times N$ matrices $\lambda_n$ $(n \in \mathbb{Z}_+)$ the following three conditions are equivalent:*

  (1) *For all $s \in \mathbb{Z}_+$*

(6.1.1)
$$\lambda_{s+d} + a_1 \lambda_{s+d-1} + \cdots + a_d \lambda_s = 0$$

422(2) $\sum_{n=0}^{\infty} \lambda_n t^n = \frac{P(t)}{Q(t)}$, where $P(t)$ is an $N \times N$ matrix polynomial with entries of degree $< d$,

(3) For all $s \in \mathbb{Z}_+$

(6.1.2) $$\lambda_s = \sum_{i=1}^{k} P_i(s)\gamma_i^s,$$

where $Q(x) = \prod_{i=1}^{k}(1 - \gamma_i x)^{m_i}$, the numbers $\gamma_i$ are distinct and $P_i(x)$ is an $N \times N$ matrix polynomial with entries of degree $< m_i$ ($i = 1\ldots k$).

Furthermore, for a sequence of complex $N \times N$ matrices $\lambda_n$ ($n \in \mathbb{Z}$) the conditions (6.1.1) and (6.1.2) for all $s \in \mathbb{Z}$ are equivalent and they imply

(6.1.3) $$\sum_{n=1}^{\infty} \lambda_{-n} t^n = -\frac{P(t^{-1})}{Q(t^{-1})}$$

*Proof.* The proof is the same as in [S] for the $N = 1$ case. □

Given a positive odd integer $\ell$, define a *Frobenius map* $R \to R^F$ on the set of rational functions $\mathbb{C}(t)$ as follows. We may write $R = c\prod_i (t - a_i)^{m_i}$ where $c, a_i \in \mathbb{C}, m_i \in \mathbb{Z}$; then we let $R^F = c^{\ell} \prod_i (t - a_i^{\ell})^{m_i}$. It is clear that this map is multiplicative (but not additive). It follows that we have

(6.1.4) $$R^F(t^{\ell}) = \prod_{\eta \in \mu_{\ell}} R(\eta t),$$

where $\mu_{\ell}$ denotes the set of all $\ell$–th roots of 1.

6.2 We turn now to the study of finite–dimensional irreducible $\widetilde{U}_{\varepsilon}$–modules. Denote by $\mathfrak{a}_+$, $\mathfrak{a}_-$, and $\mathfrak{a}_0$ the subalgebras of $\widetilde{U}_{\varepsilon}$ generated by all elements $E_{\alpha+n\delta}$ ($\alpha \in \Delta_+, n \in \mathbb{Z}$); $E_{-\alpha+n\delta}$ ($\alpha \in \Delta_+, n \in \mathbb{Z}$); and $K_{\alpha}$ ($\alpha \in P$), $E_{k\delta}^{(i)}$ ($k \in \mathbb{Z}^{\times}, i = 1, \ldots, n$) respectively. Then by the PBW theorem we obtain:

(6.2.1) $$\widetilde{U}_{\varepsilon} = \mathfrak{a}_- \otimes \mathfrak{a}_0 \otimes \mathfrak{a}_+ = \mathfrak{a}_- \otimes \mathfrak{a}_+ \otimes \mathfrak{a}_0.$$

Indeed the first equality in (6.2.1) follows from Proposition 1.7; the second equality follows from the first one using the fact that $\mathfrak{a}_+$ is generated by the $E_{\alpha_i+n\delta}$ and the relation (1.6.5b).

**Proposition 6.2.** *Let $V$ be an irreducible $\widetilde{U}_{\varepsilon}$–module. Then $V$ is finite–dimensional if and only if the following two conditions hold:*

(1) *There exists a common eigenvector for all the $K_{\alpha}$ ($\alpha \in P$) and $\psi_k^{(i)}$ ($i = 1, \ldots, n; k \in \mathbb{Z}^{\times}$).*

(2) *For each $i = 1, \ldots, n$ and each sign $\pm$ there exist $c_1^{\pm}, \ldots, c_N^{\pm} \in \mathbb{C}$, not all zero, such that one has in $V$:*

(6.2.2) $$\sum_{j=1}^{N} c_j^{\pm} E_{\pm\alpha_i + (j+s)\delta} = 0 \text{ for all } s \in \mathbb{Z}.$$

*Proof.* Assume $V$ is finite–dimensional. Then (1) is clear since all the $K_\alpha$ and $\psi_k^{(i)}$ mutually commute. Furthermore, (6.2.2) holds for $s = 0$; taking brackets of this with $E_\delta^{(i)}$ or $E_{-\delta}^{(i)}$ $|s|$ times gives (6.2.2) for all $s$ (due to (1.6.5$b$)).

Conversely, if (1) holds, then $V = \mathfrak{a}_-\mathfrak{a}_+v$ for some $v \in V$, due to the second equality of (6.2.1). Assume also that (2) holds. We write the elements of $\mathfrak{a}_\pm$ as (non–commutative) polynomials in the $E_{\pm\alpha_i+m\delta}$ ($i = 1,\ldots,n; m \in \mathbb{Z}$). Due to (6.2.2) and Proposition 1.7($c$), after bringing an element of $\mathfrak{a}_-\mathfrak{a}_+v$ to a PBW form, only finitely many root vectors $E_{\pm\alpha+m\delta}$ ($\alpha \in \Delta_+, m \in \mathbb{Z}$) appear. Since the $\ell$–th powers are scalars, we deduce that $\dim \mathfrak{a}_-\mathfrak{a}_+v < \infty$. $\square$

**Remark 6.2.** Taking bracket of both sides of (6.2.2) with $E_{\mp\alpha_i}$ and using (1.6.5$d$) we obtain:

$$(6.2.3) \qquad \sum_{j=1}^{N} c_j^\pm \widehat{\psi}_{j+s}^{(i)} = 0 \text{ for all } s \in \mathbb{Z},$$

where we let

$$(6.2.4) \qquad \widehat{\psi}_m^{(i)} = K_i^{\text{sgn}(m)}\psi_m^{(i)}.$$

Let $V$ be a $\widetilde{U}_\varepsilon$–module and let $v_\lambda \in V$ be a common eigenvector of all the $K_\alpha$ ($\alpha \in P$) and $\psi_k^{(i)}$ ($k \in \mathbb{Z}^\times, i = 1,\ldots,n$). We have:

$$(6.2.5) \qquad K_\alpha v_\lambda = \lambda_0(\alpha)v_\lambda, \text{ where } \lambda_0 : P \to \mathbb{C}^\times \text{ is a homomorphism.}$$

$$(6.2.6) \qquad \psi_k^{(i)} v_\lambda = \lambda_k^{(i)} v_\lambda \ (k \in \mathbb{Z}^\times, i = 1,\ldots,n), \text{ where } \lambda_k^{(i)} \in \mathbb{C}.$$

The collection $\lambda$ consisting of a homomorphism $\lambda_0 : P \to \mathbb{C}^\times$ and $n$ sequences $\{\lambda_k^{(i)}\}_{k \in \mathbb{Z}^\times}$ ($i = 1,\ldots,n$) is called the *weight* of $v_\lambda$, and $v_\lambda$ is called a *weight vector* of the $\widetilde{U}_\varepsilon$–module $V$. It is convenient to introduce the generating series

$$\lambda_\pm^{(i)}(t) = \lambda_0(\alpha_i)^{\pm 1}\left(1 \pm (\varepsilon^{d_i} - \varepsilon^{-d_i})\sum_{s=1}^{\infty}\lambda_{\pm s}^{(i)}t^s\right).$$

Then the weight is given by the collection $(\lambda_0, \lambda_+^{(i)}(t), \lambda_-^{(i)}(t))$. The following lemma is immediate from Remark 6.2 and Lemma 6.1.

**Lemma 6.2.** *Let $v_\lambda$ be a weight vector of a finite–dimensional $\widetilde{U}_\varepsilon$–module $V$. Then its weight $\lambda = (\lambda_0, \lambda_+^{(i)}(t), \lambda_-^{(i)}(t))$ satisfies the following two conditions ($i = 1,\ldots,n$):*

(1) $\lambda_+^{(i)}(t)$ *is a rational function such that*

$$(6.2.8) \qquad \lambda_+^{(i)}(0) = \lambda_0(\alpha_i), \quad \lambda_+^{(i)}(\infty) = \lambda_0(\alpha_i)^{-1}.$$

(2) $\lambda_-^{(i)}(t^{-1}) = \lambda_+^{(i)}(t)$. $\square$

6.3 In this subsection we study the "diagonal" $\widetilde{U}_\varepsilon$–modules.

**Definition 6.3** (a) An irreducible $\widetilde{U}_\varepsilon$–module $V$ is called *diagonal* if all the scalars $E_\alpha^\ell$ ($\alpha \in \widetilde{\Delta}^{\text{re}}$) are zero.

(b) A weight vector $v_\lambda$ of a $\widetilde{U}_\varepsilon$–module $V$ is called *singular* if $E_{\alpha+k\delta}v_\lambda = 0$ for all $\alpha \in \Delta_+, k \in \mathbb{Z}$.

**Theorem 6.3** (a) *Any two singular vectors of a diagonal $\widetilde{U}_\varepsilon$–module are proportional.*
  (b) *If two diagonal $\widetilde{U}_\varepsilon$–modules admit singular vectors, then these modules are isomorphic if and only if the weights of the singular vectors coincide.*
  (c) *A diagonal $\widetilde{U}_\varepsilon$–module is finite dimensional if and only if it admits a singular vector $v_\lambda$ and its weight $\lambda = (\lambda_0, \lambda_+^{(i)}(t), \lambda_-^{(i)}(t))$ satisfies the conditions (1) and (2) of Lemma 6.2.*

*Proof.* The proof of (a) and (b) and of the fact that a finite–dimensional diagonal $\widetilde{U}_\varepsilon$–module admits a singular vector follows from (6.2.2) in the usual way. The "only if" part of (c) follows from Lemma 6.2. Suppose now that $v_\lambda$ is a singular vector of an irreducible $\widetilde{U}_\varepsilon$–module $V$ and that $\lambda$ satisfies conditions (1) and (2) of Lemma 6.2. Then by Lemma 6.1 we have for some $d_i \in \mathbb{Z}_+$ and $a_j^{(i)} \in \mathbb{C}$

$$(6.3.1) \qquad (\widehat{\psi}_{s+d_i}^{(i)} + a_1^{(i)}\widehat{\psi}_{s+d_i-1}^{(i)} + \cdots + a_{d_i}^{(i)}\widehat{\psi}_s^{(i)})v_\lambda = 0 \text{ for all } s \in \mathbb{Z}.$$

Since $v_\lambda$ is singular, using (1.6.4d), (6.3.1) gives for each $i, j = 1, \ldots, n$ and $s \in \mathbb{Z}$:

$$(6.3.2) \qquad E_{\alpha_j+(s-1)\delta}(E_{-\alpha_i+(d_i+1)\delta} + a_1^{(i)}E_{-\alpha_i+d_i\delta} + \cdots + a_{d_i}^{(i)}E_{-\alpha_i+\delta})v_\lambda = 0.$$

In other words, the vector

$$v_i' := (E_{-\alpha_i+(d_i+1)\delta} + a_1^{(i)}E_{-\alpha_i+d_i\delta} + \cdots + a_{d_i}^{(i)}E_{-\alpha_i+\delta})v_\lambda$$

is annihilated by all the operators $E_{\alpha_j+s\delta}$ where $j = 1, \ldots, n, s \in \mathbb{Z}$. Since these elements generate $\mathfrak{a}_+$, it follows from (a) and (6.3.2) that $v_i' = 0$ for all $i = 1, \ldots, n$. Applying to the previous equality $E_{\pm\delta}^{(i)}$ $|m|$ times we get by (1.6.5b) for all $m \in \mathbb{Z}$:

$$(6.3.3) \qquad (E_{-\alpha_i+m\delta} + a_1^{(i)}E_{-\alpha_i+(m-1)\delta} + \cdots + a_{d_i}^{(i)}E_{-\alpha_i+(m-d_i+1)\delta})v_\lambda = 0.$$

Now due to the first equality in (6.2.1) we have:

$$V = \widetilde{U}_\varepsilon v_\lambda = \mathfrak{a}_- v_\lambda.$$

It follows from (6.3.3) that $\dim V < \infty$ in the same way as at the end of the proof of Proposition 6.2. $\square$

Given a finite–dimensional $\widetilde{U}_\varepsilon$–module the weight of its singular vector is call its *highest weight*.

**Corollary 6.3.** *Finite-dimensional diagonal $\widetilde{U}_\varepsilon$–modules are in one to one correspondence with $(n+1)$–tuples $(\lambda_0, R_1(t), R_2(t), \ldots, R_n(t))$, where $\lambda_0 : P \to \mathbb{C}^\times$ is a homomorphism and $R_i(t)$ are rational functions such that*

$$(6.3.4) \qquad R_i(0) = \lambda_0(\alpha_i), \quad R_i(\infty) = \lambda_0(\alpha_i)^{-1}.$$

*The highest weight associated to an $(n+1)$-tuple $(\lambda_0, R_1, \ldots, R_n)$ is $\lambda = (\lambda_0, \lambda_+^{(i)}(t), \lambda_-^{(i)}(t))$, where $\lambda_+^{(i)}(t) = \lambda_-^{(i)}(t^{-1}) = R_i(t)$.* □

**Remark 6.3.** If $\lambda = (\lambda_0, \lambda_+^{(i)}(t), \lambda_-^{(i)}(t))$ and $\lambda' = (\lambda'_0, \lambda'^{(i)}_+(t), \lambda'^{(i)}_-(t))$ are the highest weights of diagonal finite–dimensional irreducible representations, then $\lambda\lambda'$ (where the multiplication is coordinate–wise) is the highest weight of $V \otimes W \cong W \otimes V$. This follows from the formula

$$\Delta(\hat{\psi}_k) = \sum_{j=0}^{k} \hat{\psi}_j \otimes \hat{\psi}_{k-j} \mod (\mathfrak{a}_- \otimes \mathfrak{a}_+ + \mathfrak{a}_+ \otimes \mathfrak{a}_-),$$

which can be derived from [Be]. In the $\widetilde{\mathfrak{sl}_2}$ case, this is shown in [CP].

6.4 Denote by $\operatorname{Spec} \widetilde{U}_\varepsilon$ the set of all finite–dimensional irreducible representations of the algebra $\widetilde{U}_\varepsilon$. By Schur's lemma we have the canonical map

(6.4.1) $$\chi : \operatorname{Spec} \widetilde{U}_\varepsilon \longrightarrow \operatorname{Spec} Z_\varepsilon.$$

We now turn to the basic problem of calculating the image of $\chi$, which we denote by $\mathcal{F}$. We shall identify the Poisson algebraic groups $\operatorname{Spec} Z_\varepsilon$ and $\Omega$ using the isomorphism $\pi$.

Given $\sigma \in \operatorname{Spec} \widetilde{U}_\varepsilon$ we call its image $\chi(\sigma) \in \Omega$ the *central character* of the representation $\sigma$. Note that the central character of any irreducible subquotient of $\sigma_1 \otimes \sigma_2$ ($\sigma_1, \sigma_2 \in \operatorname{Spec} \widetilde{U}_\varepsilon$) is equal to the product of central characters $\chi(\sigma_1)\chi(\sigma_2)$ in $\Omega$. It follows that $\mathcal{F}$ is a subgroup of $\Omega$.

Now note that we have canonical embeddings $i_0 : \mathbb{C}(t) \to \mathbb{C}((t))$ and $i_\infty : \mathbb{C}(t) \to \mathbb{C}((t^{-1}))$ obtained by expanding a rational function in a Laurent series at $0$ and at $\infty$ respectively. Denote by $\mathbb{C}(t)_0$ the subalgebra of $\mathbb{C}(t)$ consisting of the rational functions that are regular at $0$ and $\infty$. One has: $i_0(\mathbb{C}(t)_0) \subset \mathbb{C}[[t]]$, $i_\infty(\mathbb{C}(t)_0) \subset \mathbb{C}[[t^{-1}]]$. We shall identify $\mathbb{C}(t)$ and $\mathbb{C}(t)_0$ with their images under the maps $i_0$ and $i_\infty$.

It is convenient to look at the groups $\underline{H}((t^{\pm 1}))$ by identifying them with the groups $\mathbb{C}((t^{\pm 1}))^\times \otimes_\mathbb{Z} Q^\vee$ (where $n \in \mathbb{Z}$ acts on $a \in \mathbb{C}((t^{\pm 1}))^\times$ as $a \mapsto a^n$ and on $\lambda \in Q^\vee$ as $\lambda \mapsto n\lambda$). Given a finite–dimensional representation of $\mathfrak{g}$ in a vector space $V$, an element $\sum_i a_i \otimes \lambda_i \in \underline{H}((t^{\pm 1}))$ acts on a weight vector $v_\mu \in V$ by the formula:

$$\left(\sum_i a_i \otimes \lambda_i\right) v_\mu = \prod_i a_i^{(\lambda_i|\mu)}.$$

Note that one has:

(6.4.2) $$\underline{H}[[t^{\pm 1}]] = \mathbb{C}[[t^{\pm 1}]]^\times \otimes_\mathbb{Z} Q^\vee = \mathbb{C}[[t^{\pm 1}]]^\times \otimes_\mathbb{Z} P.$$

Consider the subgroup $\tilde{H}_+ = \mathbb{C}(t)_0^\times \otimes P$ of $\underline{H}[[t]] \subset \underline{G}[[t]]$ obtained using $i_0$ and the subgroup $\tilde{H}_- = \mathbb{C}(t)_0^\times \otimes P$ of $\underline{H}[[t^{-1}]] \subset \underline{G}[[t^{-1}]]$ obtained using $i_\infty$.

Denote by $\tilde{N}_+^{\mathrm{rat}}$ the subgroup of $\underline{G}[[t]]$ generated by $h(t) \in \tilde{H}_+$ such that $h(0) = 1$, by $\exp a(t)e_\beta$ such that $a(t) \in i_0(\mathbb{C}(t)_0)$ and $\beta \in \Delta_+$, and by $\exp a(t)e_{-\beta}$ such that $a(t) \in i_0(\mathbb{C}(t)_0)$, $a(0) = 0$ and $\beta \in \Delta_+$. Similarly, denote by $\tilde{N}_-^{\mathrm{rat}}$ the subgroup of $\underline{G}[[t^{-1}]]$ generated by $h(t^{-1}) \in \tilde{H}_-$ such that $h(\infty) = 1$, by $\exp a(t^{-1})e_\beta$ such that $a(t^{-1}) \in i_\infty(\mathbb{C}(t)_0)$, $a(\infty) = 0$ and $\beta \in \Delta_+$, and by $\exp a(t^{-1})e_{-\beta}$ such that $a(t^{-1}) \in i_\infty(\mathbb{C}(t)_0)$ and $\beta \in \Delta_+$.

Let

(6.4.3) $$\Omega^{\mathrm{rat}} = \{(hu_-, h^{-1}u_+) \mid u_\pm \in \tilde{N}_\pm^{\mathrm{rat}},\ h \in H\}.$$

**Remark 6.4.** Denote by $\underline{G}((t^{\pm 1}))^{\mathrm{rat}}$ the subgroup of $\underline{G}((t^{\pm 1}))$ that consists of elements $g$ such that $\mathrm{Ad}\, g$ on $\mathfrak{g}((t^{\pm 1}))$ is in the Chevalley basis a matrix with elements in $\mathbb{C}(t)_0$. Let $\tilde{\tilde{G}}_{\mathrm{rat}} = \underline{G}((t^{-1}))^{\mathrm{rat}} \times \underline{G}((t))^{\mathrm{rat}}$. Then $\Omega^{\mathrm{rat}} = \tilde{\tilde{G}}_{\mathrm{rat}} \cap \Omega$. This follows from the results of [A–S] on Chevalley groups over semilocal rings, since the algebra $\mathbb{C}(t)_0$ is semilocal.

Introduce the following subgroup of the group $\Omega^{\mathrm{rat}}$:

$$\Omega_0^{\mathrm{rat}} = \{(a,a) \mid \mathrm{Ad}\, a \in (Ad\underline{G})(\mathbb{C}(t)_0), a(0) = hn_+, a(\infty) = h^{-1}n_-,\ \in H, n_\pm \in N_\pm.\}$$

6.5 First, we calculate the image of the "diagonal" part of the map $\chi$. This will give, in particular, the image of the set of diagonal finite–dimensional irreducible representations.

In (1.6.4) replace $u$ by $\eta t$ where $\eta \in \mu_\ell$ to obtain:

(6.5.1) $$\exp\left((q^{d_i} - q^{-d_i}) \sum_{k=1}^{\infty} E_{k\delta}^{(i)} \eta^k t^k\right) = 1 + (q^{d_i} - q^{-d_i}) \sum_{k=1}^{\infty} \psi_k^{(i)} \eta^k t^k.$$

Multiplying all equalities (6.5.1) over $\eta \in \mu_\ell$ we obtain after specializing to $\tilde{U}_\varepsilon$:

(6.5.2) $$\exp\left(\sum_{k=1}^{\infty} x_{k\delta}^{(i)} t^{k\ell}\right) = \prod_{\eta \in \mu_\ell} \left(1 + (\varepsilon^{d_i} - \varepsilon^{-d_i}) \sum_{k=1}^{\infty} \psi_k^{(i)} (\eta t)^k\right).$$

In a similar fashion we obtain

(6.5.3) $$\exp\left(-\sum_{k=1}^{\infty} x_{-k\delta}^{(i)} t^{k\ell}\right) = \prod_{\eta \in \mu_\ell} \left(1 - (\varepsilon^{d_i} - \varepsilon^{-d_i}) \sum_{k=1}^{\infty} \psi_{-k}^{(i)} (\eta t)^k\right).$$

Consider a finite–dimensional irreducible representation $\sigma$ of $\tilde{U}_\varepsilon$ in a vector space $V$ and let $v_\lambda \in V$ be a vector of weight $\lambda$. Applying both sides of (6.5.2) and (6.5.3) to $v_\lambda$, we obtain

(6.5.4) $$\exp\left(\pm \sum_{k=1}^{\infty} x_{\pm k\delta}^{(i)}(\chi(\sigma)) t^k\right) = z_{\alpha_i}^{\mp 1}(\chi(\sigma)) \lambda_\pm^{(i)}(t)^F.$$

Hence we have

(6.5.5) $$\pi_\pm^0 = \sum_{i=1}^{n} z_i^{\mp 1}(\lambda_+^{(i)}(t)^F) \otimes \omega_i^\vee.$$

by Lemma 6.2 (2). In other words we have (see (5.2.2)):

(6.5.6) $$\pi_0^{0\pm 1} \pi_\pm^0 = \sum_{i=1}^{n} \lambda_+^{(i)}(t)^F \otimes \omega_i^\vee \in \underline{H}(Z_-^0 \otimes Z_0^0 \otimes Z_+^0).$$

**Remark 6.5.** For the study of finite dimensional representations it is convenient to write the map $\pi : \operatorname{Spec} Z_\varepsilon \to \Omega$ in the form $\pi = (\pi'_-(u'), \pi'_+(u'))$ where

$$\pi'_- = \prod_{k>0} \exp(z_{\beta_k} x_{-\beta_k} e_{-\beta_k}) \prod_{k<0} \exp(-z_{\alpha_i}^{-1} x_{-k\delta} t^{-k} \omega_i^\vee) \prod_{k\leq 0} \exp(x_{-\beta_k} e_{-\beta_k}),$$

$$\pi'_+ = \prod_{k\leq 0} \exp(-z_{\beta_k} x_{\beta_k} e_{\beta_k}) \prod_{k>0} \exp(z_{\alpha_i} x_{k\delta} t^k \omega_i^\vee) \prod_{k>0} \exp(-x_{\beta_k} e_{\beta_k}).$$

In this way the factors $\pi_0^0(h)^{\pm 1}$ in (5.3.7) are incorporated into $\pi'_\pm$ (see Remark 5.2). This allows us to make the identification between the rational functions in the first and second components of $\Omega$.

Let $\tilde{H}_0$ be the subgroup of $\tilde{\Omega}_0^{\mathrm{rat}}$ consisting of elements of the form $(i_\infty(h(t)), i_0(h(t)))$, where $h(t) = \sum_i R_i(t) \otimes \omega_i^\vee$, $R_i(t) \in \mathbb{C}(t)_0$ and $R_i(0)R_i(\infty) = 1$.

**Proposition 6.5.** *The image under the map $\chi$ of the set of all diagonal irreducible finite–dimensional representations is $\tilde{H}_0$. The image of the representation with highest weight $(\lambda_0, \lambda_+^{(i)}(t), \lambda_-^{(i)}(t))$ is $(h(t), h(t))$ with $h(t) = \sum_i \lambda_+^{(i)}(t)^F \otimes \omega_i^\vee$.*

*Proof.* The proposition follows from (6.5.6) and Corollary 6.3. $\square$

We also have the following nice corollary of (6.5.4):

**Corollary 6.5.** *The rational functions $\lambda_\pm^{(i)}(t)^F$ are independent of the choice of the weight $\lambda$ of a finite–dimensional irreducible representation of $\tilde{U}_\varepsilon$.* $\square$

6.6 Recall that the definition of the real root vectors is based on the reduced expression $x = s_{j_1} \ldots s_{j_d}$. Consider the set of root vectors

(6.6.1) $$S = \{T_{i_d}^{-1} \ldots T_{i_{k+1}}^{-1} E_{i_k} \mid 0 \leq k \leq d\}.$$

It is clear that every real root vector as defined in (1.6.1) is some power of $T_x$ applied to a unique element of $S$. For each $\alpha \in \tilde{\Delta}_+^{\mathrm{re}}$ fix $w = s_{i_1} s_{i_2} \ldots s_{i_{k-1}} y$ where $1 \leq i_j \leq n$ $y \in Q^\vee$ and $\alpha_{i_k} \in \tilde{\Pi}$ so that $w(\alpha_{i_k}) = \alpha$. Define $E'_{\pm\alpha} = T_w(E_{\pm\alpha_i})$. By [Be, Proposition 6.1 and Proposition 2.3] we have for each $E_s \in S$

(6.6.2) $$E_s = P(E'_\alpha, h_{k\delta}^{(i)}), \ 1 \leq i \leq n, \ \alpha \in \tilde{\Delta}^{\mathrm{re}}$$

where P is a polynomial.

**Lemma 6.6.1.** *Let $V$ be a finite dimensional representation of $\tilde{U}_\varepsilon$. For $\alpha \in \Delta$ the elements $\{E'_{\alpha+j\delta} = T_{i_1} T_{i_2} \ldots T_{i_{k-1}} T_{\omega_{i_k}^\vee}^j E_{i_k} \mid j \in \mathbb{Z}\}$ act on $V$ in such a way that there exist $c_1, \ldots, c_N \in \mathbb{C}$, not all zero, such that one has in $V$:*

(6.6.3) $$\sum_{j=1}^N c_j^\pm E'_{\pm\alpha+(j+s)\delta} = 0 \ \textit{for all} \ s \in \mathbb{Z}.$$

*Proof.* This is proved in a similar manner to Proposition 6.2 where the element $E_{\pm\delta}^{(i)}$ is replaced by $T_{i_1} T_{i_2} \ldots T_{i_{k-1}} T_{\omega_{i_k}^\vee}^d E_{\pm\delta}^{(i_k)}$. $\square$

**Lemma 6.6.2.** *Let $V$ be a finite dimensional $\widetilde{U}_\varepsilon$–module. Then the roots vectors $\{E_{\alpha+k\langle\alpha,x\rangle\delta} \mid k \in \mathbb{Z}\}$ act in a quasipolynomial manner on $V$ (i.e. their matrix entries are finite linear combinations of functions of the form $P(k)\lambda^k$ where $P$ is a polynomial and $\lambda \in \mathbb{C}$).*

*Proof.* Through each root vector $E_\alpha \in S$ consider set $\{T_x^k E_\alpha = E_{\alpha-k\langle\alpha,x\rangle\delta} \mid k \in \mathbb{Z}\}$. These sets exhaust all possible roots. Each $E_\alpha$ for $\alpha \in S$ is some polynomial in the $E'_\alpha$ and imaginary root vectors (6.6.2). Since $W$ normalizes $P^\vee$ we can assume $T_x T_{i_1} \ldots T_{i_{k-1}} = T_{i_1} \ldots T_{i_{k-1}} T_y$ for some $y \in P^\vee$. If $y = \prod_{m=1}^n \omega_m^{\vee d_m}$ then $T_y = \prod_m T_{\omega_m^\vee}^{d_m}$ and $T_y T_{\omega_{i_k}^\vee}^d E_{i_k} = T_{\omega_{i_k}^\vee}^{d+d_{i_k}} E_{i_k}$. Therefore $T_x E'_\alpha = E'_{\alpha+k'_{x\alpha}\delta}$ for all $\alpha \in \widetilde{\Delta}^{\text{re}}$ where $k'_{x\alpha}$ and $d_{i_k} \in \mathbb{Z}$ is dependent on $x$ and $\alpha$. Using (6.6.2) and the fact that $T_x(h_k^{(i)}) = h_k^{(i)}$ we see $T_x^k(E_\alpha) = E_{\alpha-k\langle\alpha,x\rangle\delta} = P(E'_{\alpha+k'_{x\alpha}\delta}, h_{k\delta}^{(i)})$ where $k \in \mathbb{Z}$ and $P$ are independent of $\mathbb{Z}$. From here using (6.6.3) the lemma follows. $\square$

Our main result is

**Theorem 6.6.** $\mathcal{F} = \Omega_0^{\text{rat}}$.

*Proof.* First we show that $\mathcal{F} \subset \Omega_0^{\text{rat}}$. By the definition of the map $\pi$ and in view of Remark 6.5 and (6.5.6), it suffices to prove the following

**Lemma 6.6.3.** *Entries of the matrices $\pi_+^+$, $\pi_+^-$, $\pi_-^+$, $\pi_-^-$ lie in $Z_\varepsilon \otimes_\mathbb{C} \mathbb{C}(t)_0$.*

*Proof.* We shall consider the $\pi_+^+$, the proof in the remaining three cases being similar. Recall that
$$\pi_+^+ = \prod_{k \leq 0, <} \exp(x_{\beta_k} e_{\beta_k}).$$

Using the commutation formula [St, Lemma 15], we may reorder this product such that it turns into a product of expressions of the form

$$\exp(a \sum_{0 \leq j_1 \leq \cdots \leq j_s} x_{\beta_1+j_1\delta} \ldots x_{\beta_s+j_s\delta} e_{\beta_1+\cdots+\beta_s+(j_1+j_2\cdots+j_s)\delta})$$

(6.6.4)
$$= \exp(a \sum_{0 \leq j_1 \leq \cdots \leq j_s} (x_{\beta_1+j_1\delta} t^{j_1}) \ldots (x_{\beta_s+j_s\delta} t^{j_s}) e_{\beta_1+\cdots+\beta_s}),$$

where $\beta_1, \ldots, \beta_s \in \Delta_+$ and $a \in \mathbb{C}$.

Note that by Lemma 6.6.2 and Lemma 6.1 the entries of the matrices of elements $E_{\beta+k\delta}$ in a finite–dimensional representation $\sigma$ of $\widetilde{U}_\varepsilon$ are quasipolynomials in $k$ (i.e. a finite linear combination of functions of the form $P(k)\lambda^k$, where $P$ is a polynomial and $\lambda \in \mathbb{C}$). Hence the same is true for $E_{\beta+k\delta}^\ell$. If $\sigma$ is irreducible then all eigenvalues of $E_{\beta+k\delta}^\ell$ are equal to $x_{\beta+k\delta}(\chi(\sigma))$. Since their sum is the trace of $E_{\beta+k\delta}^\ell$, we conclude that $x_{\beta+k\delta}(\chi(\sigma))$ is a quasipolynomial in $k$. It follows by Lemma 6.1 that the entries of the matrices (6.6.4) lie in $\mathbb{C}(t)_0$. $\square$

In order to prove the inclusion $\mathcal{F} \supset \Omega_0^{\text{rat}}$, we make the following two observations:

(6.6.5) $$\mathcal{F} \supset \widetilde{H}_0$$

by Proposition 6.5(a), and

(6.6.6) $\quad (a,a) \in \mathcal{F} \Longrightarrow$ the connected component of
$K \cdot (a,a) \cap \Omega$ containing $(a,a)$ lies in $\mathcal{F}$.

Observation (6.6.6) follows from Proposition 4.1 and the remark in [DC–K–P3, §4.1] which says, in particular, that if $\sigma \in \operatorname{Spec} \widetilde{U}_\varepsilon$, so that $\pi(\sigma) \in \mathcal{F}$, then the whole symplectic leaf of $\pi(\sigma)$ lies in $\mathcal{F}$.

Since $\Omega_0^{\mathrm{rat}}$ is generated by $\tilde{H}_0$ and the 1–parameter root subgroups as descibed in §6.4, in view of (6.6.5) it suffices to show that for a given $a \in t\mathbb{C}(t)_0$ there exists $\beta, \gamma \in \mathbb{C}[t, t^{-1}]$ and $h \in \mathbb{C}(t)_0^\times$ such that

$$\begin{pmatrix} h^{-1} & 0 \\ 0 & h \end{pmatrix} \begin{pmatrix} 1 & 0 \\ \beta & 1 \end{pmatrix} \begin{pmatrix} h & 0 \\ 0 & h^{-1} \end{pmatrix} \begin{pmatrix} 1 & 0 \\ \gamma & 1 \end{pmatrix} = \begin{pmatrix} 1 & 0 \\ a & 1 \end{pmatrix}.$$

Equivalently, given $a \in t\mathbb{C}(t)_0$ we wish to find $\beta, \gamma \in \mathbb{C}[t, t^{-1}]$ and $h \in \mathbb{C}(t)_0$ such that

(6.6.7) $$a = h^2 \beta + \gamma.$$

This is straightforward. $\square$

6.7 For $\widetilde{U}_\varepsilon = U_\varepsilon(\widetilde{\mathfrak{sl}_2})$ Theorem 6.6 means the following. Let $g(t) = -\sum_{k=0}^\infty z_\alpha x_{\alpha+k\delta} t^k$, $h(t) = z_\alpha \exp(\sum_{k=1}^\infty x_{k\delta} t^k) = z_\alpha (1 + (\varepsilon - \varepsilon^{-1}) \sum_{k=1}^\infty \psi_k t^k)^F$, $f(t) = -\sum_{k=1}^\infty x_{-\alpha+k\delta} t^k$. Using Lemma 6.1 it follows that $\mathcal{F} = \{(A(t), A(t))\}$ where

(6.7.1) $$A(t) = \begin{pmatrix} 1 & g(t) \\ 0 & 1 \end{pmatrix} \begin{pmatrix} h(t)^{1/2} & 0 \\ 0 & h(t)^{-1/2} \end{pmatrix} \begin{pmatrix} 1 & 0 \\ f(t) & 1 \end{pmatrix}.$$

We shall now calculate the image under $\chi$ of an evaluation representation. We let $E = E_1$, $F = F_1$, $K = K_\alpha$. Recall [DC–K] that the center of $U_\varepsilon(\mathfrak{sl}_2)$ is generated by the elements $x = (\varepsilon - \varepsilon^{-1})^\ell E^\ell$, $y = (\varepsilon - \varepsilon^{-1})^\ell F^\ell$, $z_1 = K_\omega^\ell$, $c = (\varepsilon - \varepsilon^{-1})^2 FE + K\varepsilon + K^{-1}\varepsilon^{-1}$. Let $z = z_1^2$.

Recall that due to Jimbo [J2] there exists for $a \in \mathbb{C}^\times$ an "evaluation" homomorphism $ev_a : U_q(\widetilde{\mathfrak{sl}_2}) \to U_q(\mathfrak{sl}_2)$ given by

$$ev_a(E_{-\alpha+k\delta}) = (aq^{-1})^k K^k F,$$
$$ev_a(E_{\alpha+k\delta}) = (aq^{-1})^k E K^k.$$

One can show that:

$$ev_a(1 + (q - q^{-1}) \sum_{k=1}^\infty \psi_i t^k) = \frac{a^2 q^{-4}(t^2 - a^{-1} q^2 c_q t + a^{-2} q^4)}{(1 - aq^{-3} Kt)(1 - aq^{-1} Kt)}$$

where $c_q = (q - q^{-1})^2 FE + qK + q^{-1}K^{-1}$.

Consider a finite–dimensional irreducible representation $\sigma$ of $U_\varepsilon(\mathfrak{sl}_2)$ with the prescribed values of central elements $x$, $y$, $z$, and $c$. Then $\tilde{\sigma}_a := \sigma \circ ev_a \in \operatorname{Spec} U_\varepsilon(\widetilde{\mathfrak{sl}_2})$. We have $\chi(\tilde{\sigma}_a) = (g(t), g(t))$, where

$$(6.7.2) \qquad g(t) = \begin{pmatrix} 1 & -\frac{xz}{1-a^\ell zt} \\ 0 & 1 \end{pmatrix} \begin{pmatrix} h(t)^{1/2} & 0 \\ 0 & h(t)^{-1/2} \end{pmatrix} \begin{pmatrix} 1 & 0 \\ -\frac{zya^\ell t}{1-a^\ell zt} & 1 \end{pmatrix}$$

Here $h(t) = \frac{(1-a^\ell c_1^\ell t)(1-a^\ell c_1^{-\ell} t)}{(1-a^\ell zt)^2}$, where $c = c_1 + c_1^{-1}$. Note that if $x = 0$ then $h(t) = \frac{1-a^\ell z^{-1} t}{1-a^\ell zt}$.

It follows that taking central characters of tensor products of all $\tilde{\sigma}_a$ where $\sigma$ is a diagonal representation of $U_\varepsilon(\mathfrak{sl}_2)$, we get the whole subgroup $\tilde{H}_0$ of $\Omega_0^{\text{rat}}$. Furthermore, taking all $\tilde{\sigma}_a$ with $\sigma$ having central character $x, y = 0, z = 1$ (resp. $x = 0$, $y$, $z = 1$), we get all matrices

$$\begin{pmatrix} 1 & \frac{\alpha}{1+\beta t} \\ 0 & 1 \end{pmatrix} \text{ and } \begin{pmatrix} 1 & 0 \\ \frac{\alpha t}{1+\beta t} & 1 \end{pmatrix}.$$

Since these matrices together with $\tilde{H}_0$ generate the group $\Omega_0^{\text{rat}}$ [AS], we obtain

**Proposition 6.7.** *The image under $\chi$ of the set of all irreducible subquotients of tensor products of evaluation representations of $U_\varepsilon(\widetilde{\mathfrak{sl}_2})$ coincides with $\Omega_0^{\text{rat}}$.* □

Department of Mathematics, Harvard University, Cambridge, MA 02138
*E-mail address*: beck@math.harvard.edu

Department of Mathematics, Massachusetts Institute of Technology, Cambridge, MA 02139
*E-mail address*: kac@math.mit.edu